\documentclass[preprint,aps,showpacs]{revtex4} 
\usepackage{graphicx} 
\usepackage{dcolumn} 
\usepackage{amsmath} 
\usepackage{bm} 
\usepackage{epsfig} 
\begin{document} 
 
\preprint{HEP/123-qed} 
 
\title[Dynamic Rearrangements]{Dynamic Rearrangements and 
Packing Regimes in  Randomly Deposited Two-Dimensional Granular Beds} 
\author{I. Bratberg } 
\altaffiliation[Temporary address:]{ Fysisk Institutt, Universitetet i Oslo, Postboks 1048 Blindern, N-0316 Oslo, Norway}
\affiliation{Department of Telecommunications,  
Norwegian University of Science and Technology,  
N-7491 Trondheim, 
Norway.} 
 
\author{F. Radjai} 
\affiliation{LMGC, CNRS-Universit\'e Montpellier II,  
Place Eugène Bataillon,  
34095 Montpellier cedex, France.} 
 
\author{A. Hansen} 
\affiliation{Department of Physics,  
Norwegian University of Science and Technology, 
N-7491 Trondheim, 
Norway.} 
 
\date{\today} 
 
\begin{abstract} 
We study the structural properties of two-dimensional 
granular packings prepared by 
random deposition from a source line.  
We consider a class of random ballistic deposition models  
based on single-particle  
relaxation rules controlled by a critical angle, and we  
show that these local  
rules can be formulated as rolling friction in the framework of  
dynamic methods for the simulation of granular materials.  
We find that a packing prepared by    
random deposition models is generically unstable, and  
undergoes dynamic rearrangements. As a result, the dynamic  
method leads systematically to a higher solid fraction than the  
geometrical model for the same critical angle. We characterize  
the structure of the packings generated by both methods in terms  
of solid fraction, contact connectivity and anisotropy.  
Our analysis provides evidence for  four packing regimes as 
a function of solid fraction, the mechanisms of packing growth being 
different in each regime.   

\end{abstract} 
 
 
\maketitle 
 
\section{Introduction} 
 
 
Random ballistic deposition (RBD) is a well-known method for  
layer-by-layer construction of random  packings of hard  
particles such as granular beds and colloidal  
aggregates\cite{houi,jullien1}. This method is  
based on a simple and intuitive procedure.   
The particles (mainly monodisperse spheres or disks) are  
allowed to fall sequentially along randomly  
positioned vertical lines over a horizontal substrate. Upon   
contact with the substrate or the first (already deposited)  
particle, the particle either sticks or is    
further moved to a more favorable position according  
to a relaxation (or restructuring) rule. The RBD method  
can be efficiently implemented in a computer code  
for generating very large two- and three-dimensional  
packings. Elaborate large-scale simulations based on this approach  
have been used to investigate the    
geometrical properties of random packings (packing regimes,  
distribution functions, growth, etc)\cite{jullien2,jullien3}.  
 
 
It is obvious that the random deposition of particles   
can also be  simulated by means of dynamic methods, such as molecular  
dynamics\cite{allen,luding} and contact dynamics\cite{moreau,jean,radjai},    
in the spirit of  a real experiment where the grains are poured  
into a box.  Such simulations require, however, substantially  
more computation time\cite{schaefer}. This difficulty has been  
inhibiting enough to discourage for a long time systematic  
investigation of  
deposited beds following dynamic methods. But, the situation  
is far better today due to the fast increase of  
available computer power and memory during the last decade.  
There is now a considerable scope for dynamic simulations  
that can be exploited  
in order to study a number of highly interesting  
issues in the field of random packings. 
 
 
The objective of the present paper is to apply   
a contact dynamics algorithm  to investigate   
randomly-deposited granular beds in a two-dimensional geometry.   
The geometrical texture (coordination number,  
solid fraction, \ldots) of a granular bed   
depends on several physical parameters (particle properties,  
contact interactions, inertia of deposited particles)  
which can be tuned in a dynamic simulation in order to characterize  
the impact of each parameter on the texture. 
We propose here  an original approach which allows  
to bring out some interesting features of granular beds in  
comparison to RBD models. We consider a generalized  
RBD model in which the relaxation of the falling particle  
upon contact with the substrate is controlled by the  
direction of the contact normal  
$\theta$ (a single angle in 2D)\cite{bdm}. Depending on whether $\theta$  
is below or above a critical angle $\theta_c$,  the particle either  
simply sticks or is allowed to rotate until it reaches a  
local minimum position or forms a new contact below $\theta_c$.   
Hereafter, we refer to this model  
as the CA (Critical Angle) model. The central feature of this model  
is that it allows to control the solid fraction $\rho$ by varying  
the critical angle\cite{bdm}.  
The approach we propose consists in performing dynamic simulations  
of random ballistic deposition {\em as closely as possible to  
the CA model}.  
 
This requires that we transcribe the  
the above geometrical relaxation rule into a  
contact law which is  reduced to the geometrical rule    
for the random deposition process. We first show that this   
requirement is met if the particles interact through  
a {\em rolling friction} law (similar to the Coulomb sliding   
friction) in which a contact torque is mobilized to resist  
relative rotation of two particles.  
We implement this law within a contact dynamics (CD) algorithm.  
Then, we perform two series of simulations both  
with the CA model and the CD method. In the first series,  
we use the granular beds prepared according to  the CA procedure   
as initial configuration for a CD simulation. We show  
that, although local stability (sticking due to rolling  
friction or, alternatively, particles supported by  
two underlying contacts) is  
fulfilled for each particle added to the bed, the latter still undergoes  
collective rearrangements leading to a higher solid fraction.  
This implies that  a granular packing prepared by geometrical  
rules is globally unstable.  
We study the extent of dynamic rearrangements and 
the structural properties of the CA and CD packings as  
a function of the critical angle.  
   
In a second series of simulations, we characterize the packings  
in terms of the average coordination number, structural anisotropy  
and contact connectivity as a function of the solid fraction.  
We show that the trends are globally similar for  the CD method  
and the CA model (the same packing regimes are observed). We  
distinguish several packing regimes where different mechanisms  
(screening, chaining, branching, piling, jamming and ordering)  
are active and control the packing fraction.       
 
\section{Numerical approach}   
 
\subsection{Critical-angle model} 
 
\begin{figure}[h] 
\includegraphics[width=12cm]{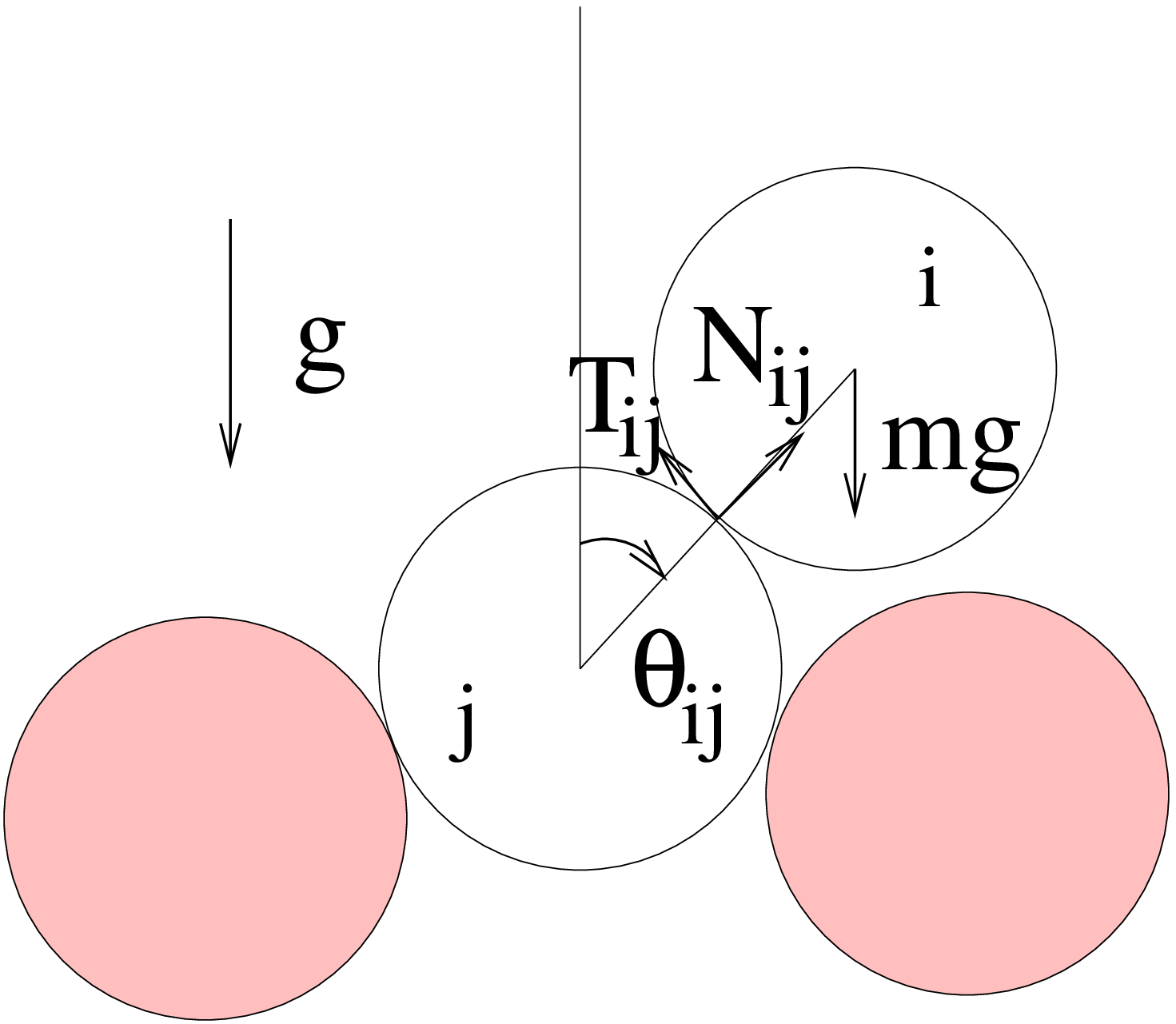} 
\caption{\label{fig1} Geometry of a contact formed by a falling 
particle i with a particle j from the substrate. The contact 
angle $\theta_{ij}$ is measured from the vertical.} 
\end{figure}

 
Figure~\ref{fig1} shows the geometry of a contact 
formed by a falling particle i  
with a particle j of the substrate. 
The two particle centers define a line inclined  
at an angle $\theta_{ij}$ to the vertical.  
For disks, $\theta_{ij}$  is also the  
direction of the contact normal ${\bm n}_{ij}$, unit   
vector directed from the center of particle j to the center of  
particle i. For brevity, we  
will  refer to $\theta_{ij}$ as the ``contact direction''.  
 
The CA model is defined as follows\cite{bdm}.  
If $|\theta_{ij}|$ is below a critical angle  
$\theta_c$, defined in the range between $0$ (vertical direction)  
and $\pi/2$ (horizontal direction), particle  
i simply freezes by sticking to particle j. On the other hand,  
if $|\theta_{ij}|$ exceeds $\theta_c$, particle i is allowed  
to rotate around particle j until a second  contact is formed  
with another particle k of the substrate. Then, there are three possible  
alternatives: 1) If $|\theta_{ik}| < \theta_c$, particle i freezes  
as in the first case; 2) If the new position of particle i is  
a local minimum position, again particle i freezes; 3)  
If neither of the two latter conditions is fulfilled, particle i is  
again allowed to rotate around particle k until a new contact is formed  
with another particle of the bed and the three alternatives are  
examined again with this new contact. This procedure is iterated until  
particle i is stabilized either by sticking or by reaching a local minimum  
position.  
 
Two limits are of particular interest.    
When  $\theta_c =90^\circ$, all particles stick irreversibly to  
the substrate where ever they land. This limit  
corresponds to the random sequential adsoption 
model\cite{meakin85,vold,lubachevsky,tang}.      
When $\theta_c = 0^\circ$, all particles relax and  
the model is reduced to the steepest  
descent model \cite{visscher,jullien1}.   
In \cite{meakin85}, the solid fraction  
was  found to be $\rho_{min}=0.3568\pm 0.0001$ in the no-restructuring  
limit.  The solid fraction  for the steepest descent model is expected to  
come close to $\rho_{max}=0.906$ corresponding to  a triangular  
packing.   
However, to achieve a structure with long-range  
ordering, the initial conditions are very  
important. In practice,   
the bottom line must initially be covered by an array of  
contiguous disks. Otherwise,   
simulations using the steepest descent algorithm have shown  
that the solid fraction will not exceed  
$\rho = 0.82$ which is the characteristic density   
of 2D monodisperse random close packing (RCP) where long-range  
order is broken by defects in the packing \cite{meakin85,tjueaatte}.  
 
\begin{figure}[h] 
\includegraphics[width=12cm]{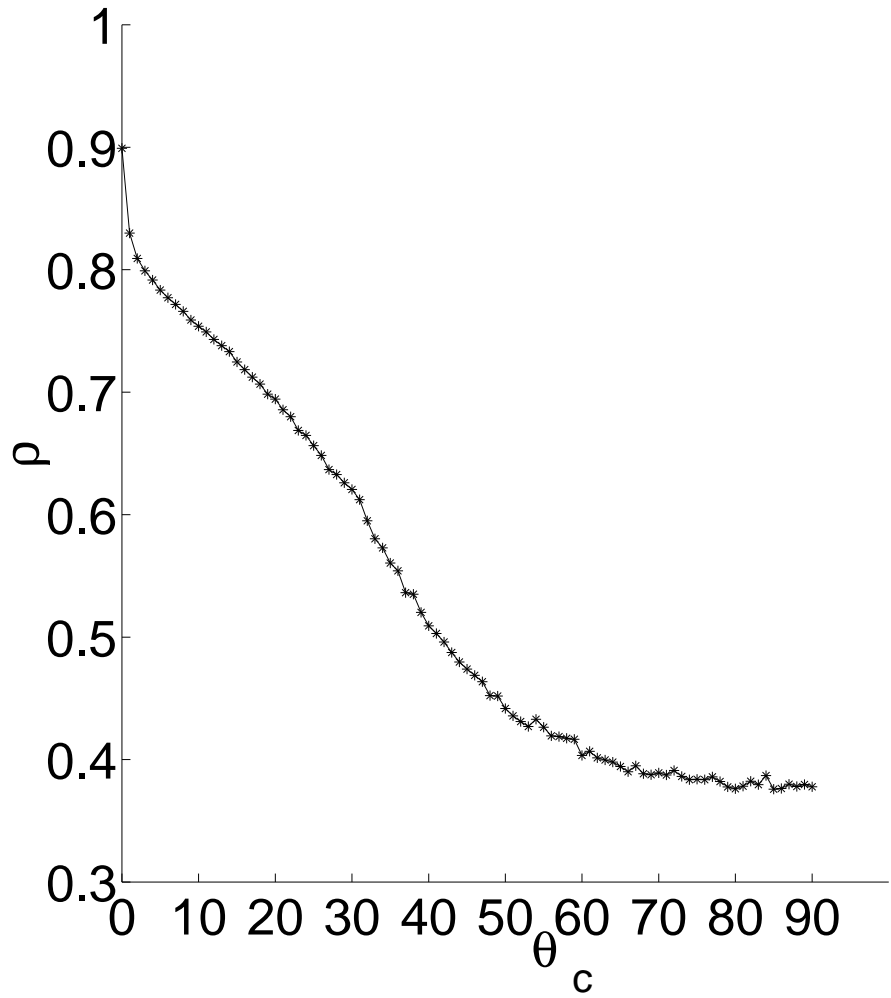} 
\caption{\label{fig2} Solid fraction $\rho$ as a 
function of critical angle $\theta_c$ (in degrees) 
for the CA model.} 
\end{figure}

In the CA model, the solid fraction $\rho$ of the granular bed  
is a function of the critical angle $\theta_c$  
as shown in Fig.~\ref{fig2}. In this figure, the  
solid fraction for each of the angles $\theta_c=0,1,2, \cdots,90$  
is an average over $30$ independent runs.  The bottom line  
was covered by a layer of 32 contiguous particles and 1000 particles  
were deposited in each run.  In order to avoid wall effects, periodic  
boundary conditions were implemented in the horizontal direction.  
Let us note that CA simulations are possible at much larger  
scales. Nevertheless, we restrict here the size and the number of  
CA simulations to those reasonably accessible to  
dynamic simulations since the results will   
be compared between these two methods below.  
Figure~\ref{fig2} shows that the solid fraction decreases  
monotonously from  
$\rho_{max}$ to $\rho_{min}$  as $\theta_c$ is increased from $0$ to  
$90^\circ$. One can distinguish several  
regimes on this curve which will be discussed  
below in connection with dynamic simulations.        
 
The CA model is a geometrical model that meets  
the no-overlap condition between hard particles but    
involves a number of physical approximations about the stability  
of the packing and its growth. By nature, this model neglects  
inertia effects. The substrate is frozen and the relaxation step  
involves only the deposited particle. Moreover, the two stability criteria  
(local minimum position and sticking condition)  
for the deposited particle have a {\em local} nature. In other words,  
the model assumes that the whole packing remains in  
static equilibrium as long as all  particles are sequentially  
stabilized by either of these conditions.  
 
In order to examine the validity of these assumptions,   
the approach followed in this paper is to perform dynamic  
simulations as closely as possible  
to the CA model and to compare the resulting packings.  
This implies that the particles should be released  
sequentially and they should hit the granular bed with negligibly small  
inertia. Moreover, upon contact with the substrate, the falling  
particle should  dynamically  
stick or roll down, depending on the value of the contact angle   
with respect to the critical angle, until one of the two   
stability conditions is fulfilled. There is no  
difficulty in tuning the inertia in a dynamic simulation.  
But, we need to define a dynamic version of the  
relaxation rule.  
 
\subsection{Rolling friction} 
 
\begin{figure}[h] 
\includegraphics[width=12cm]{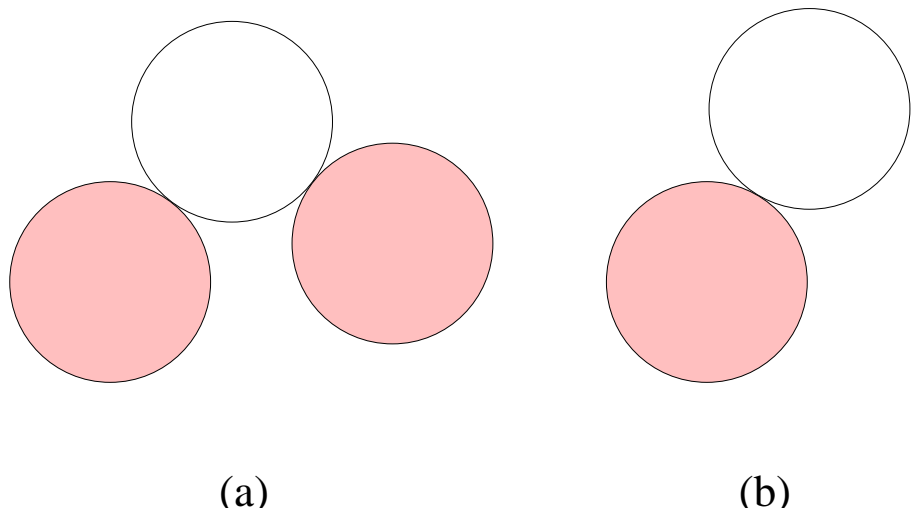} 
\caption{\label{fig3} The two local stability conditions in the CA model: 
(a) local minimum position; (b) sticking.} 
\end{figure}

The dynamic content of stability due to a  
local minimum position is clear.  
The weight of a particle can obviously be balanced by the reaction  
forces exerted by two underlying particles; see Fig.~\ref{fig3}(a). But  
the sticking condition requires both a contact force $\bm F_{ij}$ and a  
``contact torque'' $M_{ij}$ so as to counterbalance respectively  
the weight $m_i g$ of the  
deposited particle and its moment with respect to the contact point.  
Figure~\ref{fig3}(b) illustrates this condition. The balance  
equations are 
\begin{eqnarray} 
{\bm F}_{ij} + m_i {\bm g} &=& 0, \\ 
m_i g r_i \sin \theta + M_{ij} &=& 0, 
\label{eqn1}  
\end{eqnarray} 
where $r_i$ is the particle radius and ${\bm g}$ is the gravity.   
 
Let $N_{ij}$ and $T_{ij}$ be the components of  
the reaction force ${\bm F}_{ij}$  
along and perpendicular to the contact normal ${\bm n}_{ij}$.  
From equations~\ref{eqn1} one gets  
\begin{eqnarray}  
N_{ij} &=& m_i g \cos \theta, \\ 
T_{ij} &=& m_i g \sin \theta, \\  
M_{ij} &=& r_i N_{ij} \tan \theta. 
\label{eqn2} 
\end{eqnarray} 
The normal force $N_{ij}$ is positive (as it should) as long as  
$-\pi /2 <\theta < \pi /2$ (the angles are measured from  
the vertical).    
On the other hand, the relative sliding  
is inhibited if $|T_{ij}|/N_{ij} < \mu_s$, where $\mu_s$ is the  
coefficient of (sliding) friction (or equivalently,  
$\theta < \theta_s$, where $\theta_s = \tan^{-1} \mu_s$ is the  
angle of friction).  
 
Now, if we require that particle i rolls only if    
$\theta \geq \theta_r$,   
then from equations~\ref{eqn2} we arrive at the following  
no-rolling condition:  
\begin{equation} 
\frac{|M_{ij}|}{r_i N_{ij}} < \tan \theta_r = \mu_r, 
\label{eqn3}  
\end{equation} 
where $\mu_r$ is a coefficient of {\em rolling friction}.  
Let us further assume that $M_{ij}$ remains equal  
to its threshold value $\pm \mu_r r_i N_{ij}$ when rolling occurs. This  
condition is similar to the sliding  
condition $T_{ij} = \pm \mu_s N_{ij}$.  
 
\begin{figure}[h] 
\includegraphics[width=12cm]{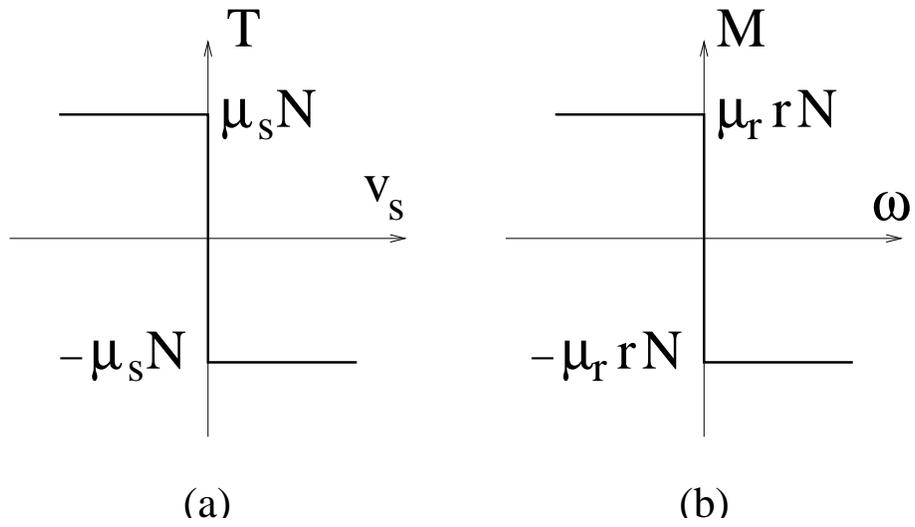} 
\caption{\label{fig4} The graphs representing (a) sliding friction 
law and (b) rolling friction law; see text.} 
\end{figure}

The rolling friction law, as defined here, and the more  
familiar Coulomb (sliding)  
friction law are shown in Fig.~\ref{fig4} in the form 
of graphs\cite{jean,radjai1}.  
The rolling friction law   
relates the relative rotation velocity  
$\omega_{ij} = \omega_i - \omega_j$ of the two  
particles to the contact torque $M_{ij}$, whereas the sliding  
friction law relates the sliding velocity $v^s_{ij}$ to the tangential  
force $T_{ij}$.   
In fact, although for the sake of clarity  
we derived the condition~\ref{eqn3} by considering the particular  
case of a deposited particle touching a particle of the bed,  
the application of rolling friction to a contact between two arbitrary  
particles in a packing is rather straightforward when formulated in the  
form of the graphs shown in Fig.~\ref{fig4}.  
The torque transmitted through a  
contact to a particle in static equilibrium, for example, is the torque  
necessary to balance the sum of all force moments and other  
contact torques acting on the particle in the same way  as the  
mobilized torque $M$  in equations ~\ref{eqn3}   
counterbalances the moment $r mg \sin \theta$ of the particle  
weight.

The prescription of rolling friction in a dynamic method  
follows the same steps as the sliding friction. The relation  
between the contact torque  and the relative rotation  
velocity (Fig.~\ref{fig4}(a))  
can {\em not} be represented as a motived function. Hence,  
in the framework  
of the molecular dynamics method, based on explicit integration of the   
equations of motion, this friction law has  to be replaced by an  
approximate {\em function}\cite{schaefer,jean,radjai3}.  
This ``regularization'' of the friction law is  
not necessary within the contact dynamics method which was employed   
for the present investigation\cite{moreau,jean,radjai}.  
 
Using either of these dynamic methods,  
the sticking condition can be achieved if  
$\mu_s$ is set to infinity (no sliding for no contact direction)  
and the angle of rolling  
friction $\theta_r$ is interpreted as the critical angle $\theta_c$.  
This was implemented in our contact dynamics simulations.  
Alternatively,  
one may set $\mu_r$ to infinity (no rolling for no contact direction)  
using the angle of sliding  
friction $\theta_s$ as the critical angle. It is also possible   
to use a combination of these two conditions. These conditions  
are not equivalent, but we will not discuss the  
differences in this paper.      
In all cases, the condition of sticking upon collision  
requires also a zero coefficient of restitution.

To summarize, the following conditions allow to perform  
a dynamic simulation of random particle deposition  
in close analogy with the CA model: 
\begin{enumerate} 
\item $\theta_r = \theta_c$ 
\item No sliding ($\theta_s = 90^\circ$)  
\item Weak inertia 
\item Zero coefficient of restitution 
\end{enumerate} 
The important difference is that, while in the CA model all degrees  
of freedom in the substrate are   
kinematically frozen, in our dynamic simulations  
only contact sliding is frozen by  
setting $\theta_s = 90^\circ$.  
All other degrees of freedom are active   
and the rolling friction governs all contacts: the  
contact between the deposited particle and the bed, as well as  
all contacts in the bed.       
Since the particles are  
not frozen in the granular bed, sequential particle deposition  
may thus lead to   
rearrangements in the granular bed.

\subsection{Simulation parameters} 
The contact dynamics (CD) simulations involve a  
number of parameters  which should be adjusted so as to 
minimize inertia effects without loosing   
numerical efficiency. The largest inertia are produced  
by the largest head-on  velocity $v_{max}$ between colliding particles.  
Let $\Delta t$  be the time step. The contact force due to inertia  
produced by a collision is $m v_{max} / \Delta t$. This force  
should be compared to the weight $mg$ of one particle.  
Hence, the  dimensionless parameter characterizing the ratio of inertia to  
weights is 
\begin{equation} 
\alpha = \frac{v_{max}}{g \Delta t} 
\label{eqn4} 
\end{equation}       
 
The influence of $\alpha$ on the solid fraction and restructuring  
is an interesting subject in itself, but it will not be  
investigated in this paper. As emphasized previously, the focus here   
is put on those effects (equilibrium states, rearrangements)  
that arise from the {\em geometrical configuration}.  
Hence, we should  use a low value of $\alpha$. However,  
lower values of  $\alpha$ mean slower simulations. Hopefully,  
the framework of the CD method allows  
for large time steps $\Delta t$ up to the limitations related to the  
procedure  of contact detection. On the other hand, the value of $v_{max}$  
can be imposed for the falling particles, but further relaxation  
inside the packing may produce large impact velocities. In particular,  
at low solid fractions, where large voids exist in the bed,  
the free fall of a particle over distances compared to the  
system height $H \simeq 60 r$ can give rise to impact forces  
far larger than the weight of a column of particles of the  
same height.  
 
In order to avoid such strong uncontrolled inertia,  
we implemented  a ``velocity barrier'' trick that  
limits the particle velocities to    
$v_{max} = 0,3$ ms$^{-1}$. With this choice, we can use  
a time step as large as $\Delta t = 0.003$ s. Then, setting  
$g=100$ ms$^{-2}$, we get $\alpha=1$. This means that  
the largest impact force is just equal to the weight of  
a single particle. This choice is both reasonable and   
compatible with numerical efficiency.

\section{Dynamic rearrangements} 
 
\subsection{Stability of CA packings}  
How stable are the granular beds  
prepared by means of the CA model? We have  
seen that the CD method,   
equipped with rolling friction together with suitable values of    
the parameters reducing inertia effects, 
meets the {\em single-particle} stability  
criteria of the CA model in the course of deposition, 
namely 1) the sticking condition  
as a function of the rolling friction angle $\theta_r$   
(identified with the critical angle $\theta_c$),       
and 2) the local minimum position where the weight of a particle  
is balanced by the reaction forces at the two underlying contacts.   
Now, if we start a CD simulation using a packing constructed  
according to the CA model as the initial configuration, then  
one of the two following alternatives may occur. If the  
single-particle stability criteria used in the course of deposition   
provide a sufficient condition for the    
{\em global} stability of the packing when the deposition   
is over, then the packing will remain in static equilibrium  
and the calculated forces will exactly balance all particles.  
Otherwise, the packing will be unstable and the CD simulations   
allow to calculate the particle rearrangements until a  
relaxed stable configuration is obtained.  
 
Our simulation data confirm rather the second alternative for nearly  
all values of the critical angle. One example is shown  
in figures~\ref{fig5}(a) and~\ref{fig5}(b)   
for a packing of 500 particles with $\theta_c=40^\circ$.  
  The rearrangements occur in the whole bed, but they are much more hindered  
in the bulk than in the vicinity of the free surface.  
For this reason, the displacements appear mostly in  
the uppermost layers. The relaxed configuration has a larger  
packing friction. The solid fraction is still larger when the CD  
simulation is performed by random  
deposition of the same sequence of particles (as in the CA simulation)   
for the same value of $\theta_c$ (and the same boundary conditions),  
instead of using the CA configuration  
as the initial condition. The resulting packing is shown  
in Fig.~\ref{fig5}(c). In this latter case, the hindering effect  
related to the bulk density, which was active  
in the case (b), disappears since  
the CD rearrangements occur naturally in the course of deposition for  
each deposited particle. This means that the degree of instability  
of the CA packing shown in Fig.~\ref{fig5}(a)  
is more keenly reflected in the increase $\Delta \rho$ of the  
solid fraction from (a) to (c) than  
from (a) to (b).          
  
Figure~\ref{fig6} displays the solid fraction $\rho$  
as a function of $\theta_c$ for packings prepared by CD sequential  
random deposition. The solid fraction for each value of $\theta_c$   
is an average over $10$ independent CD runs.   
The CD simulations were performed for $40$ different angels.    
The curve of  
$\rho$ as a function of $\theta_c$ corresponds thus to 410  
CD simulations of 1000 particles. Figure~\ref{fig6} shows that,  
as expected, the solid fraction is every where larger for  
the CD method than for the CA model, except at $\theta_c=0$ where   
a dynamic method requires an exceptionally high precision to reach a  
perfect triangular packing. Indeed, in this limit, tiny  
fluctuations in particle positions around a particle  
due to numerical overlaps are exponentially amplified  
in space as a result of long-range order\cite{berryman}.   
Disregarding this pathological limit, the difference $\Delta \rho$  
is negligibly small for $\theta_c < 20^\circ$ and  
$\theta_c > 80^\circ$. The largest variation $\Delta \rho$  
of the solid fraction,  
representing the largest dynamic rearrangements in the packing,   
occurs at $\theta_c \simeq 50^\circ$, where $\rho$ increases from  
0.45 for the CA packing to 0.6 for the CD packing.

\subsection{Influence on the packing structure} 
The coordination number $z$ (average  
number of contact particles around a particle),  
shown in Fig.~\ref{fig7} as a function of $\theta_c$,  
follows the same trends as the solid fraction. It is systematically  
larger in a CD packing than in the corresponding CA packing  
(for the same value of $\theta_c$) except  
in the $\theta_c = 0$ limit. A coordination number close   
to $2$ reflects the predominance of particle ``chains'' in a highly  
porous packing. The coordination number  
increases from 2 to 3 due to ``branching'', and from  
3 to 4 due to a growing interplay of chains.  
The increase of $z$ beyond 4 requires long-range 
ordering\cite{gervois}. This  
transition occurs only in the CA packing where the numerical  
precision is less stringent than in dynamic simulations.   
 
Due to dynamic rearrangements in CD random  
deposition, the CA and CD packings show also very different aspects  
as to the directional order of the contact network.  
The nonuniform distribution of contact directions can be characterized  
by means of the fabric tensor $\bm \phi$ defined from  
contact normals  
${\bm n}^k = (\sin \theta^k, \cos \theta^k)$ 
by\cite{satake,radjai2}  
\begin{equation} 
\phi_{\alpha \beta} = \frac{1}{N_c}  
\sum_{k=1}^{N_c} n^k_\alpha n^k_\beta, 
\label{eqn5} 
\end{equation} 
where $N_c$ is the total number of contacts, and $n^k_\alpha$  
(resp. $n^k_\beta$) is the $\alpha$ (resp. $\beta$) component of  
the contact normal k. When the probability distribution  
function $p(\theta)$ of contact directions is known,  
the fabric tensor can be calculated from the integral  
\begin{equation} 
\phi_{\alpha \beta} =  
\int_{-\pi/2}^{\pi/2} n^k_\alpha n^k_\beta p(\theta) \ d \theta.  
\label{eqn6} 
\end{equation}  
 
By construction, we have $\phi_1 + \phi_2 = tr({\bm \phi}) =1$,  
where $\phi_1$ and $\phi_2$ are the eigenvalues.  
The mean contact direction in the packing is given by the major  
principal direction $\theta_f$ of $\bm \phi$.  
The structural anisotropy of the packing is represented by  
$a=2(\phi_1 - \phi_2)$. The factor 2 is introduced in order  
to identify this value of $a$ with that appearing naturally  
in a sinusoidal distribution    
$p(\theta) = (1/\pi)\{1 + a \cos 2(\theta - \theta_f)\}$\cite{rothenburg}.   
 
Figure~\ref{fig8} shows the anisotropy of our granular beds  
as a function of $\theta_c$. We see that the anisotropy of  
CD packings is systematically below that of CA packings. This is  
mainly because collective rearrangements tend to destroy    
columnar structures in a CD packing.  
In both cases, the anisotropy comes very close to zero  
for $\theta_c=0$. This effect is mainly related to the presence  
of a great number of particles with 5 and 6 contacts. In fact,  
using Eq.~\ref{eqn5}, it can be shown that the anisotropy  
for the set of 6 contacts around a particle is zero, and for  
a set of 5 contacts around a particle can not exceed a threshold  
imposed by steric exclusions.      

The largest anisotropy  
in the CD packings is reached for $\theta_c=90^\circ$, whereas  
the anisotropy of the CA packing passes through a maximum  
at $\theta_c \simeq 50^\circ$.                 
 The anisotropy  can be estimated analytically  
at $\theta_c=90^\circ$ where  
the packing growth is governed by sticking. Since the  
particles are released at random horizontal positions, the probability  
that a particle sticks at a contact angle $\theta$ (with respect  
to the vertical) is $p(\theta) = \frac{1}{2} \cos \theta$. Note 
that the latter is a normalized  
probability density function over the range $[-\pi/2, \pi/2[$.  
Using Eq.~\ref{eqn6} with this expression for $p(\theta)$, we find  
$a (\theta_c = 90^\circ) = 2/3$ which is consistent with both  
CD and CA results at $\theta_c= 90^\circ$ shown in Fig.~\ref{fig8}.      
 
Since $p(\theta)$ is an even function of $\theta$ ($p(\theta) = p(-\theta)$), 
the major principal direction of the fabric tensor is  
vertical ($\theta_f = 0$). However, this is only a consequence of symmetry  
and it does not imply that the distribution $p(\theta)$ is peaked 
on $\theta=0$. In fact, within each of the half-intervals  
$[-\pi/2, 0]$ and $[0,\pi/2]$, the contacts have preferred directions.  
This can be seen in one example of $p(\theta)$ for $\theta_c=0$  
shown in Fig.~\ref{fig9}. We observe a local maximum at $\theta=0$, but there 
are local maxima also in each of the half-intervals.      
In order to extract the useful information about the direction of 
contacts, one can   
calculate the fabric tensor  $\bm \phi$  by restricting  
the definition to one of the above two half-intervals.   
 
The major principal direction $\theta_f$ for  
the interval $[0,\pi/2]$ as a function of $\theta_c$  
is displayed in Fig.~\ref{fig10}. At $\theta_c=0$ and $\theta_c=90^\circ$  
both methods give the same direction, but every where else  
 the contacts are more biased to the  
horizontal direction in CD simulations compared to CA simulations.   
This is an indication that the collective rearrangements reorganize   
contact directions. As for the anisotropy, the value of $\theta_f$ can  
be calculated analytically in the 
limit $\theta_c=90^\circ$ over the interval  
$[0,\pi/2]$ from the fabric tensor.  
The distribution function normalized over this interval is given  
by $p(\theta) = \cos \theta$ and the integral in Eq.~\ref{eqn6} is  
calculated over the same interval. We find  
$\theta_f(\theta_c=90) \simeq 32^\circ$, in agreement with the  
simulation result shown in Fig.\ref{fig10}.       
 
The existence of a local minimum in the evolution of $\theta_f$ for  
the CA model at $\theta_c \simeq 30^\circ$ or the changing of  
behavior in the CD  curve at the same point can be understood as  
a consequence of competition between sticking and  
rolling. As  $\theta_c$ is increased from zero, an  
increasing number of particles stick to the substrate at an angle  
in the interval $[0, \theta_c]$. These include both the ones sticking  
to the bed upon the first collision (whose number increases  
as $\int_0^{\theta_c} \cos \theta \ d \theta = \sin \theta_c$)  
and a number of the relaxed particles. On average, this subset of  
contacts tends to decrease $\theta_f$ as long as $\theta_c$ is not  
too large. This explains the decrease of $\theta_f$ from $45^\circ$   
at $\theta_c = 0$ to $\simeq 28^\circ$ at $\theta_c \simeq 30^\circ$.      
But, the inclination of the contacts to the vertical increases  
at the same time following the increase of $\theta_c$. This trend  
dominates clearly the evolution of $\theta_f$ beyond   
$\theta_c = 30^\circ$.  
  
\section{Packing regimes} 
 
The results presented in the last section show that, for a given  
value of the critical angle $\theta_c$, the solid   
fraction and the structure of the packing differ considerably  
from the CA model to the CD approach (excepted in the two limits  
of very loose and very dense packings). We attributed these differences   
to dynamic restructuring in CD packings as the particles are 
added to the substrate. 
However, in this section, we will show that the 
structure of a CA packing 
is quite similar to that of a CD packing if they are compared at the 
same solid fraction $\rho$ (and thus, for different critical angles). 
This means that the structural properties of CA packings are 
quite realistic (close to CD packings) when they are 
considered as a function of the solid fraction rather than  
the critical angle.          
 
\subsection{Fabric} 
 
Figure~\ref{fig11} shows the coordination number $z$  
as a function of solid fraction $\rho$ for CA and CD packings.  
In both cases, $z$ increases with $\rho$. The two curves almost  
collapse for $\rho < 0.6$. For $0.6 < \rho < 0.8$, the CD packings  
show only a slightly larger coordination number than the CA packings. 
For $0.8 < \rho$,  the CD packings  
show a slightly lower coordination number than the CA packings. 
The solid fraction $\rho=0.8$ corresponds  to $z \simeq 4$ in both 
methods. 
 
The anisotropy $a$ of the packings as a function of $\rho$   
is displayed in Fig.\ref{fig12}. The anisotropy decreases as a function  
of $\rho$ for both methods except in the loosest CA packings where it  
increases a bit with $\rho$ and passes through  a peak before decreasing.  
The relatively low rate of decrease in the range $\rho <  0.6$ suggests   
that sticking is the dominant mechanism of growth in this regime, whereas  
rolling (or relaxation) is far more efficient in the subsequent range.  
 
Figure~\ref{fig13} shows the major principal direction $\theta_f$  
of the fabric tensor restricted to the interval  $[0,\pi/2]$ (as defined  
in the last section) as a function of $\rho$. In CD packings, the direction  
$\theta_f$, representing the average direction of contact normals  
in the interval $[0,\pi/2]$, increases quite slowly for $\rho <  0.6$  
and much faster beyond $0.6$. In CA packings,  
the anisotropy decreases for $\rho <  0.6$, passes through a minimum   
at $\rho \simeq 0.6$ and increases in the subsequent range. However, let us  
remark that, as for the anisotropy, the difference in the  
value of $\theta_f$  between the two methods as a function of $\rho$  
is quite small  as compared to the differences as a function of  
the critical angle (see Fig.~\ref{fig10}).

\subsection{Connectivity} 
 
The coordination number $z$ is an average over all  
particles in a packing. But, the number of contact neighbors $q$   
varies in a packing from particle to particle. In a monodisperse  
pile, $q$ can vary from 1 to 6. This ``connectivity disorder''  
characterizes the disposition of the particles as ``nodes'' of  
the contact network. The connectivity of a packing is given by  
the fraction $P_q$ of particles having $q$ contact neighbors.  
$P_1$ corresponds to the ``dead ends'' of particle chains.  
The larger $P_1$, the stronger ``screening'' (the dead ends did not grow  
because they were screened by faster growing structures).   
$P_2$ and $P_3$ are  related to chaining and branching, respectively.  
$P_4$ corresponds to ``piling'', i.e. a natural situation where  
a particle is supported by two underlying particles and supports  
two others. $P_5$ and $P_6$ define  ``jammed'' and ordered  
configurations.  
 
Figure~\ref{fig14} shows the connectivity numbers $P_q$  
as a function of solid fraction for $q$ varying from 1 to 6. 
The trends are globally 
 similar in CA and CD packings and the differences for the two methods   
are quite small. All connectivity numbers vary monotonously  
with solid fraction except $P_3$ which first increases   
to reach a maximum at $\rho \simeq 0.7$ and decreases  
rapidly afterwards. In the range $\rho < 0.6$, $P_3$ and $P_4$  
increase at the expense of $P_1$ and $P_2$ which decrease.  
$P_5$ and $P_6$ begin to increase significantly only at  
$\rho \simeq 0.7$ and $\rho \simeq 0.8$, respectively.  
 
The connectivity diagram $P_q$ is shown for  
three different solid fractions in Fig.\ref{fig15}.  
The largest connectivity number is 2 for $\rho<0.6$, 3 for  
$0.6 <\rho< 0.7$, and $4$ for $0.7 <\rho$.    
Interestingly, the screening effect is more important in  
CD simulations (the CD curve for $P_1$ stands above the corresponding  
CA curve). Chaining ($P_2$), branching ($P_3$) and ordering  ($P_6$) are  
slightly less important in   
CD simulations, while piling ($P_4$) and jamming ($P_5$) 
are enhanced.

 The above data show that the morphology of a CA packing is  
very close to that of a CD packing at the same solid fraction.  
Both methods suggest four packing regimes characterized 
by the properties of the packing structure as a function of solid  
fraction: 
\begin{enumerate} 
\item[(a)] $\rho<0.6$: This regime corresponds to  loose  
random packings characterized by chaining (i.e. $P_2$ is the 
largest connectivity number), branching ($P_3$ increases as a 
function of $\rho$ and becomes 
dominant at $\rho = 0.6$) and screening ($P_1$ is large).        
 
\item[(b)] $0.6<\rho<0.7$: This is the regime of  
moderate random packings characterized by the largest 
value of $P_3$ (branching) at the expense of chaining 
($P_2$) which decreases rapidly as a function of $\rho$.    
 
\item[(c)] $0.7<\rho<0.8$: This regime corresponds to  
dense close packings where piling is the main mechanism 
of growth and $P_4$ is larger than other connectivity numbers.   
 
\item[(d)] $0.8<\rho$: This is the well-known 
dense ordered packing regime \cite{tjueaatte} characterized 
by $z> 4$.   
 
\end{enumerate} 
 
The transition densities ($0.6$, $0.7$ and $0.8$) appearing in this 
classification are approximate values. A more refined 
evaluation of these specific densities requires a considerably more 
computation time and a deeper insight into the mechanisms at play 
during the packing growth.   
 
\section{Conclusion} 

We investigated the structure of a class of 
randomly deposited 
granular packings whose density is controlled by a 
geometrical parameter, referred to as the critical angle. We used both a 
random ballistic deposition model with simple relaxation 
rules (the CA model), and 
a contact dynamics algorithm (the CD method) that  
incorporates those relaxation rules through a rolling friction law. 
The CD approach naturally leads to stable packings following 
dynamic rearrangements  
while in the CA model the packing is kinematically frozen after 
each single-particle relaxation. 

The following results were shown by means of extensive simulations: 
\begin{enumerate}
\item[1)] The packings prepared according to the CA model are generically 
unstable. When fed into the CD algorithm as initial configuration, 
the CA packings undergo dynamic rearrangements. As a 
consequence, the solid 
fraction is larger in CD packings than in CA packings for the 
same critical angle (implemented as the angle of rolling friction in the 
framework of the contact dynamics method).   
\item[2)] The structural properties (anisotropy, connectivity) 
are quite comparable in CA and CD packings for the same 
solid fraction, even though significant differences were 
observed in packing anisotropies.      
\item[3)] Both methods reveal four packing regimes as a function of solid 
fraction, the prevailing mechanism of growth being different in each 
regime.               
\end{enumerate}

An important outcome of this work is to show 
that the dynamic rearrangements 
are quite weak in the very loose and very dense limits 
where the structural properties are nearly the same. This means that, 
the random sequential adsorption model 
(irreversible sticking without relaxation), leading 
to very loose packings,   
and the steepest descent model (no sticking), leading to very 
dense packings,  can be used with confidence in these two 
limits. 

The CD simulations reported in this work were meant to 
keep as close as possible 
to the CA model in order to perform comparable 
calculations with both methods. There is much more to be learned 
about the structure of particle packs generated by the CD 
method (or equivalently, molecular dynamics method). The influence 
of inertia and polydispersity on the observed packing regimes is 
currently under investigation. The shear resistance of deposited 
beds (e.g. in a biaxial compression) as a function of 
solid fraction is another important issue which we would like to address 
in near future.

\section{Acknowledgments} 
We thank Knut J\o rgen M\aa l\o y , Anders Malthe S\o rensen,  
Espen Gjettestuen and Lothar Brendel for fruitfull discussions. 
This work had it's financial support by the Research  
Council of Norway (NFR) through a Strategical University Program.   
We are also grateful to the CNRS and NFR for support through  
the Franco-Norwegian PICS program, Grant No. 753.

 
\newpage 
\newpage 
\newpage 
 
\newpage 
\newpage 
\begin{figure}[h] 
\includegraphics[width=6cm]{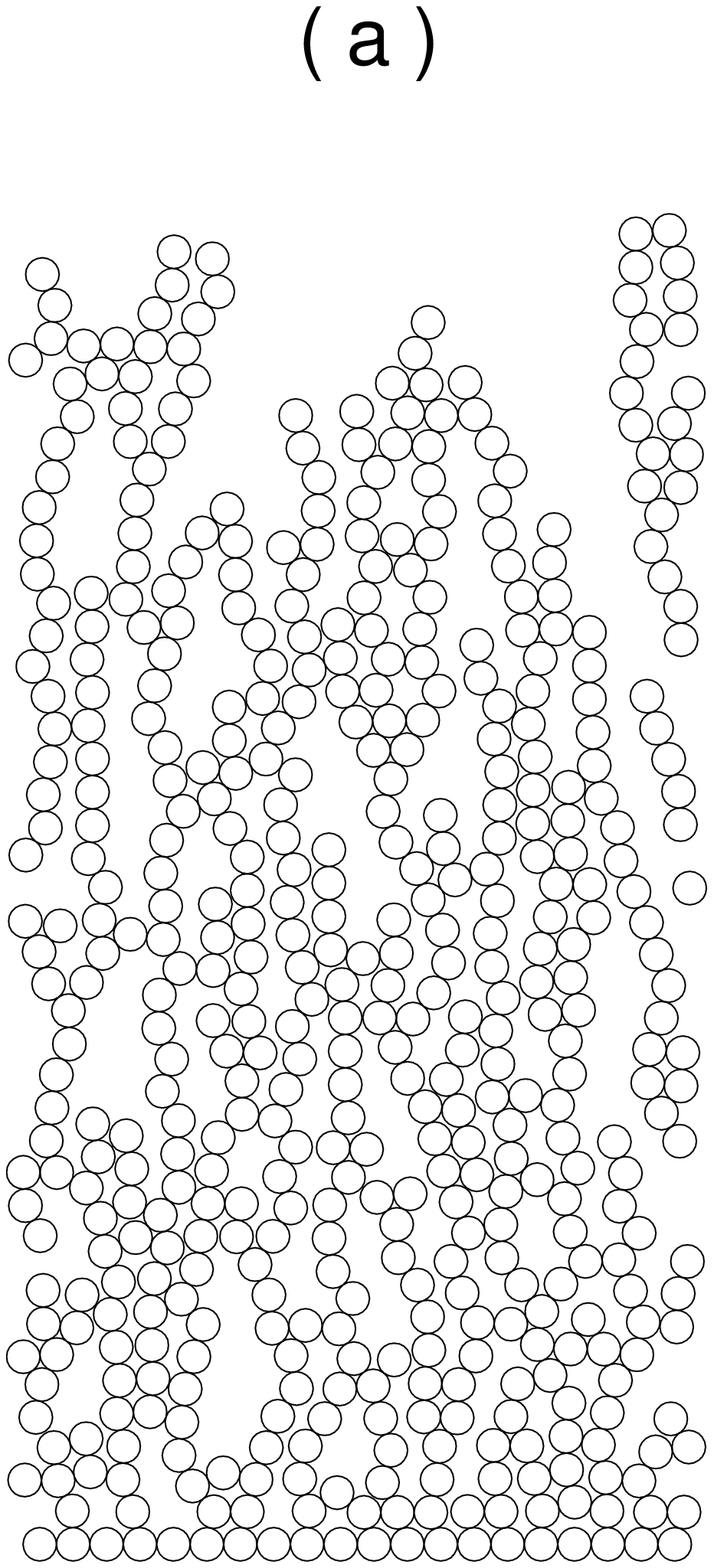}
\end{figure} 
\newpage 
\begin{figure}[h] 
\includegraphics[width=6cm]{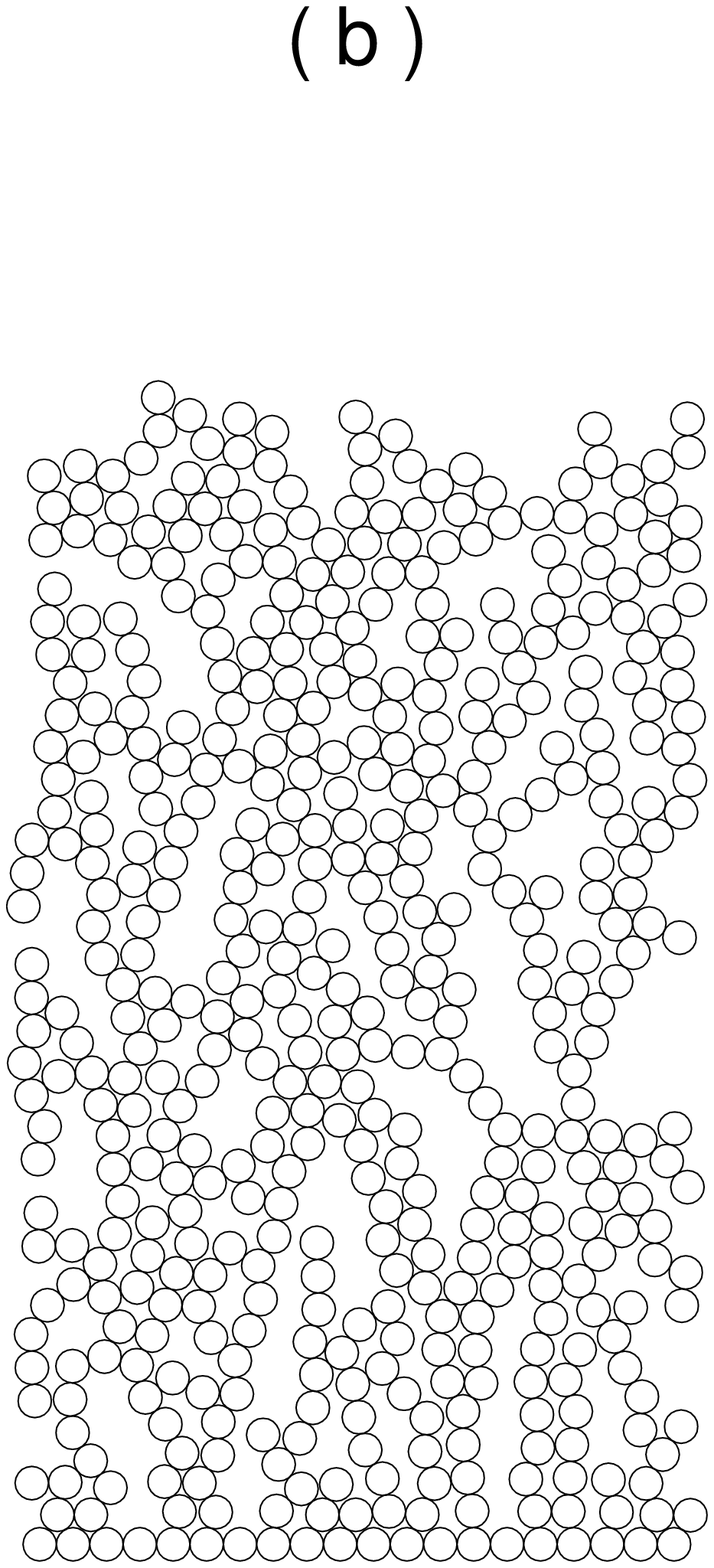}
\end{figure} 
\newpage 
\begin{figure}[h] 
\includegraphics[width=6cm]{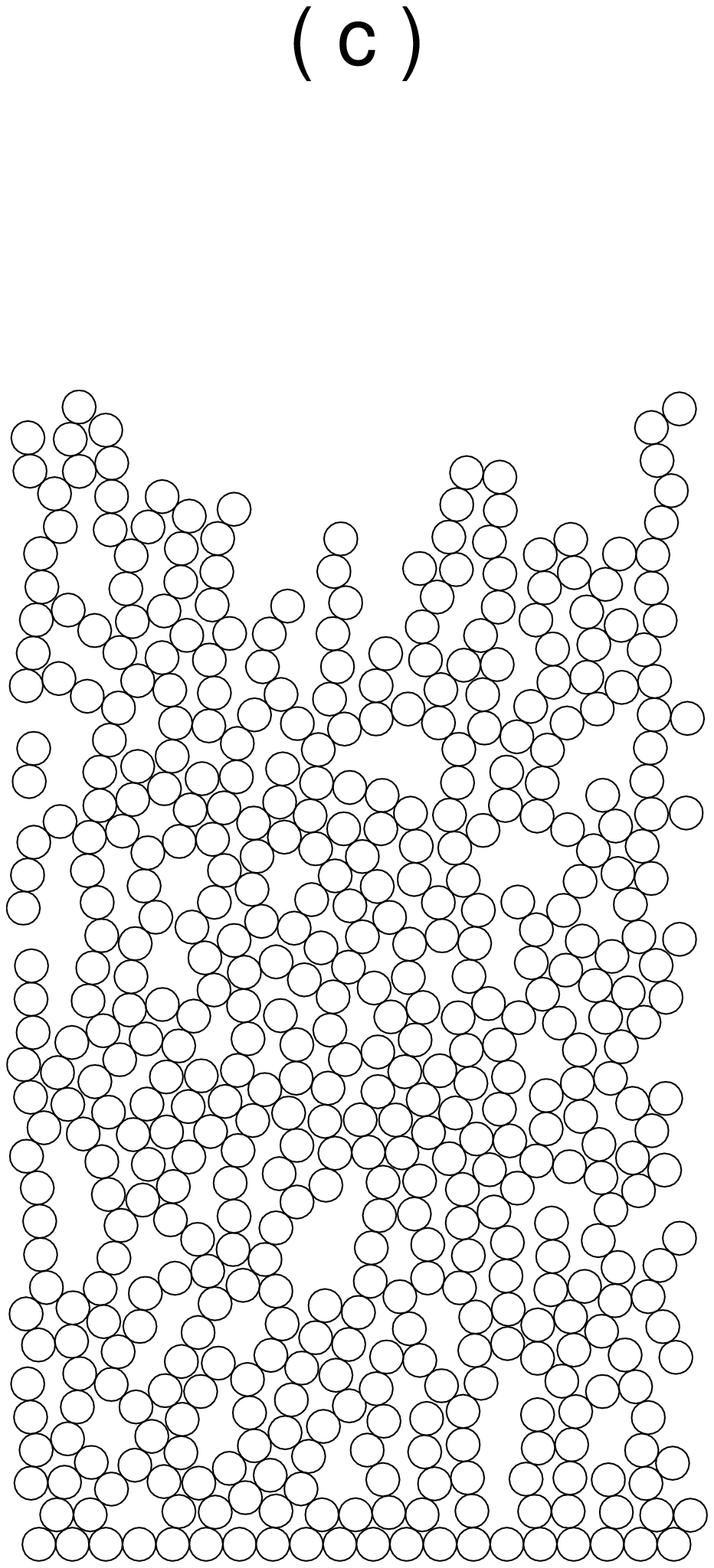} 
\caption{\label{fig5} (a) An example of a CA packing; (b) 
The static packing obtained by the CD method starting with the 
CA packing in (a) as initial condition; (c) 
CD packing obtained by using the same 
sequence of falling particles as in (a).} 
\end{figure} 
\newpage 
\begin{figure}[h] 
\includegraphics[width=12cm]{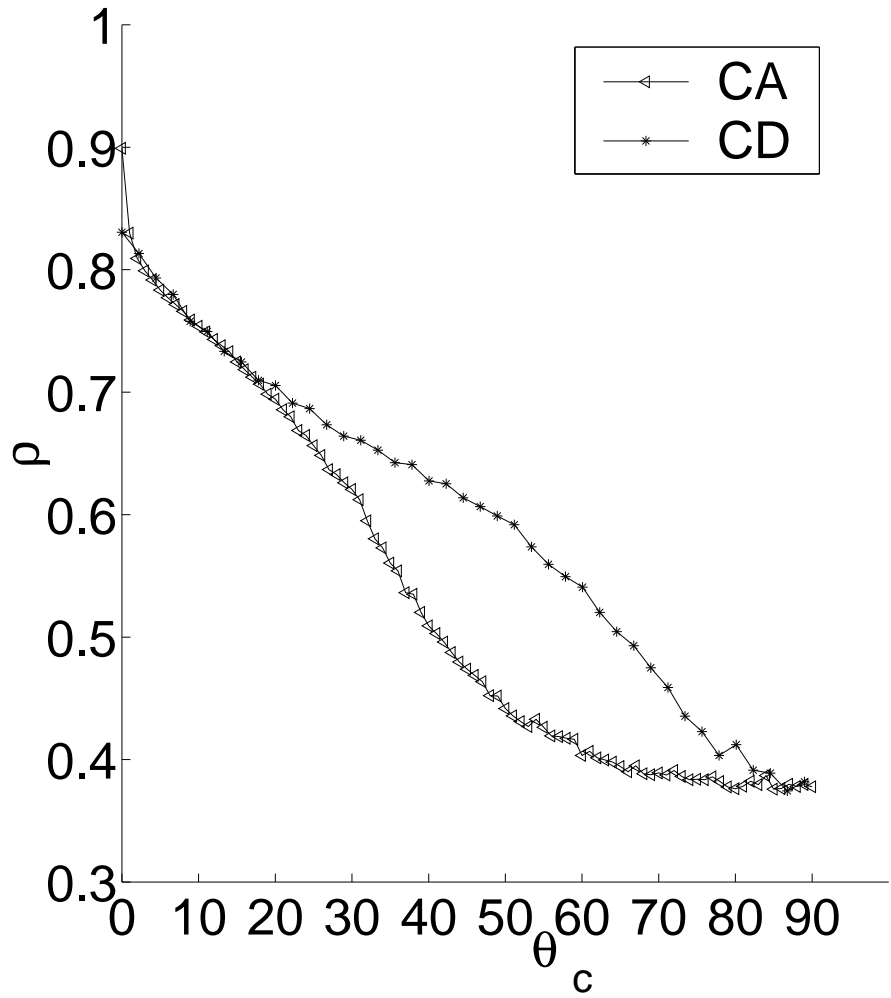} 
\caption{\label{fig6} Solid fraction as a function 
of critical angle (in degrees) for the CD and CA models.} 
\end{figure} 
\newpage 
\begin{figure}[h] 
\includegraphics[width=12cm]{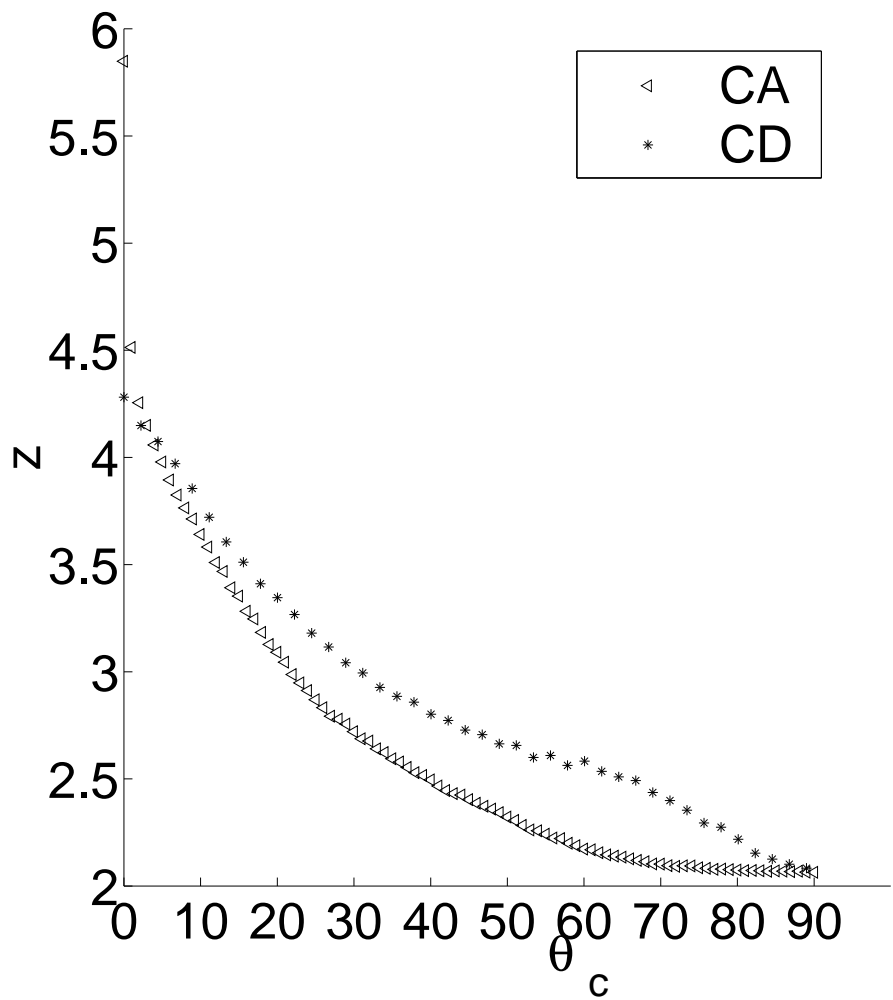} 
\caption{\label{fig7} Coordination numbers $z$ 
as a function of critical angle (in degrees).} 
\end{figure} 
 
\newpage 
\begin{figure}[h] 
\includegraphics[width=12cm]{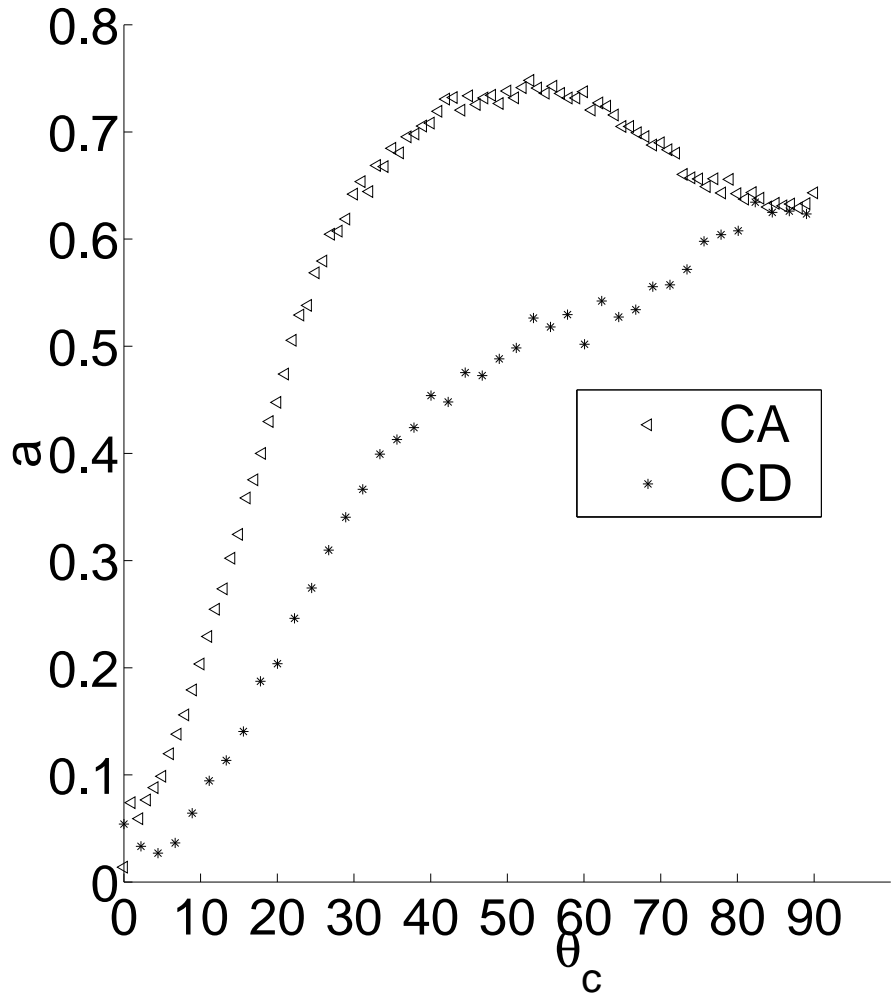} 
\caption{\label{fig8} Anisotropies $a$ as a function of 
critical angle (in degrees).} 
\end{figure} 
\newpage 
\begin{figure}[h] 
\includegraphics[width=12cm]{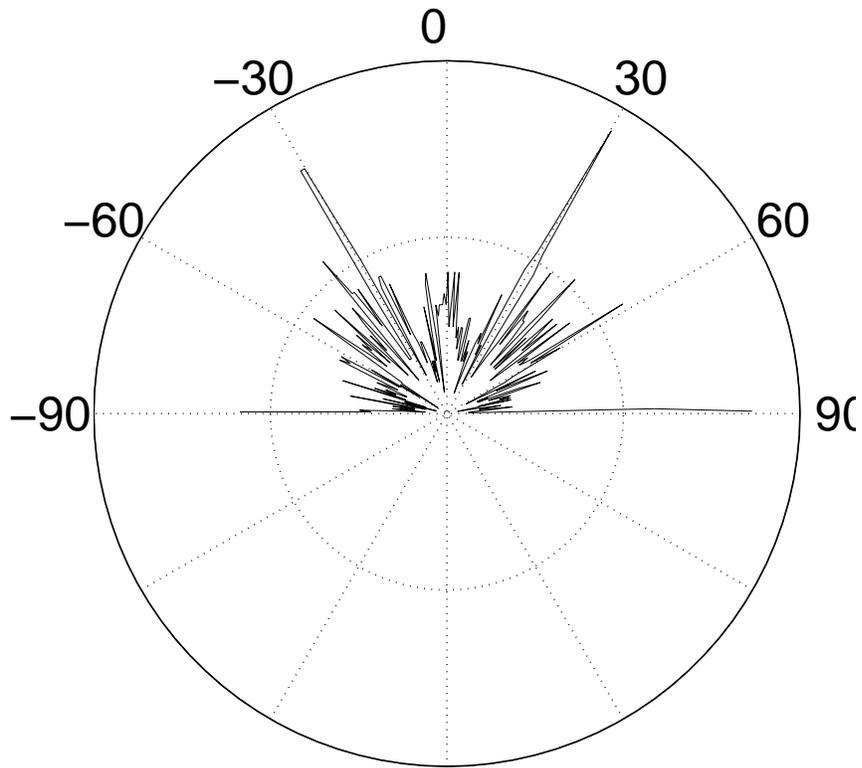} 
\caption{\label{fig9} Polar diagram of the distribution 
$p(\theta)$ of contact angles in a CD packing with $\theta_c=0$. 
The zero angle refers to the vertical direction. The angles 
are in degrees.} 
\end{figure} 
\newpage 
\begin{figure}[h] 
\includegraphics[width=12cm]{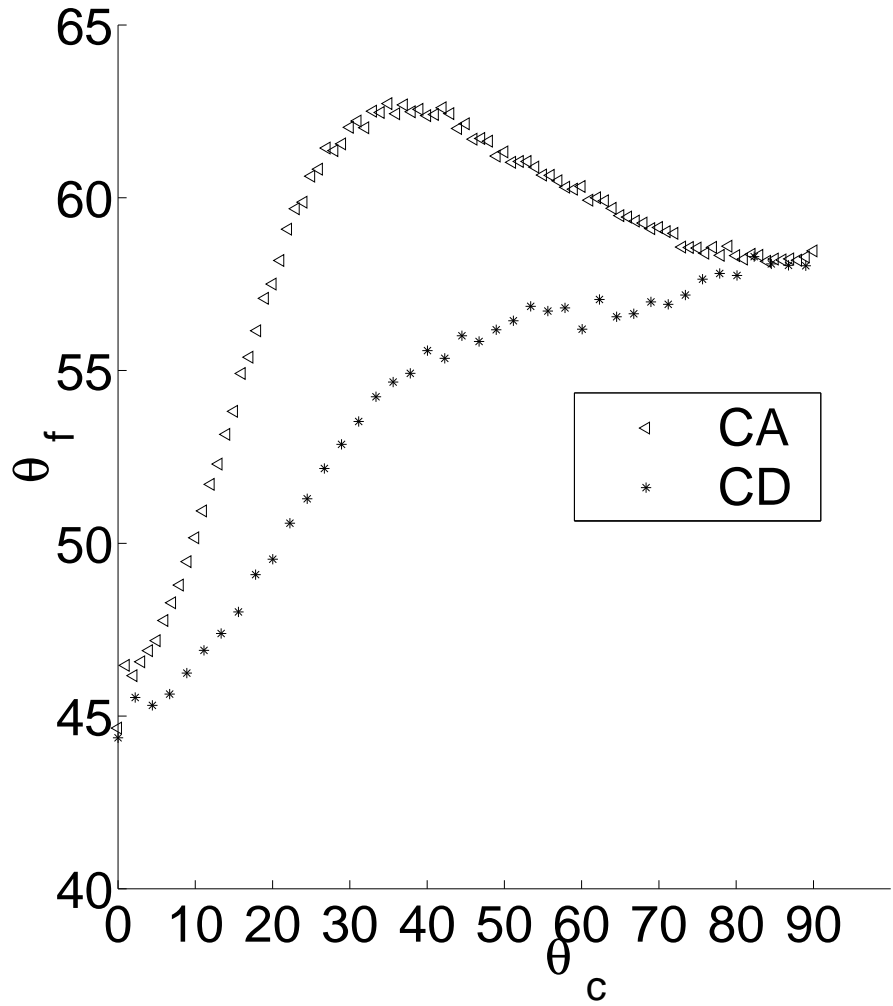} 
\caption{\label{fig10} The major principal directions  
$\theta_f$ of the fabric tensor as a function of  
critical angle. The angles are in degrees.} 
\end{figure} 
\newpage 
\begin{figure}[h] 
\includegraphics[width=12cm]{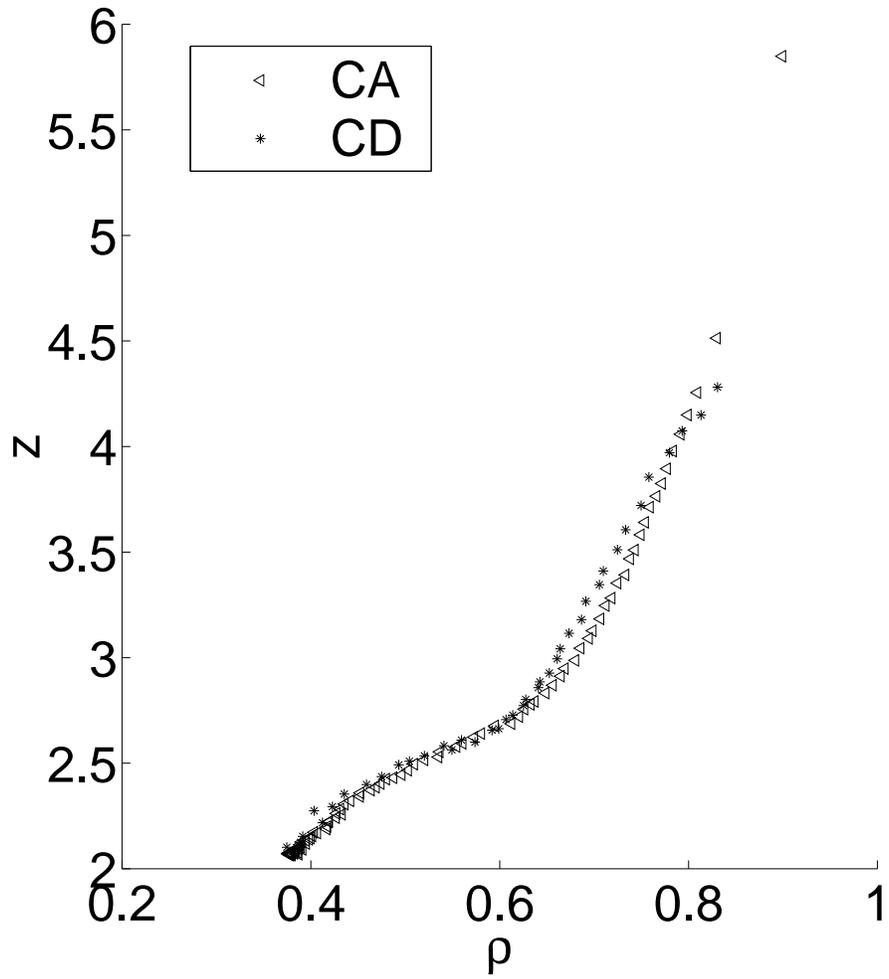} 
\caption{\label{fig11} Coordination numbers as a function of solid 
fraction $\rho$.} 
\end{figure} 
\newpage 
\begin{figure}[h] 
\includegraphics[width=12cm]{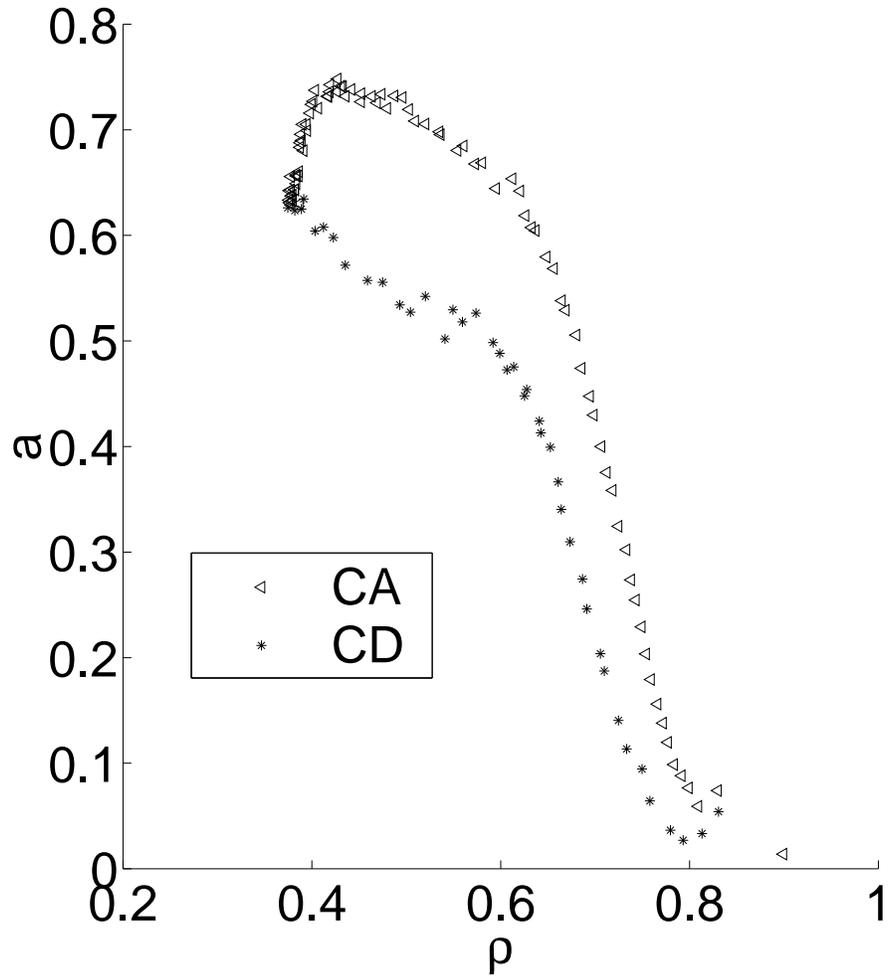} 
\caption{\label{fig12} Anisotropies $a$ as a function of 
solid fraction.} 
\end{figure} 
\newpage 
\begin{figure}[h] 
\includegraphics[width=12cm]{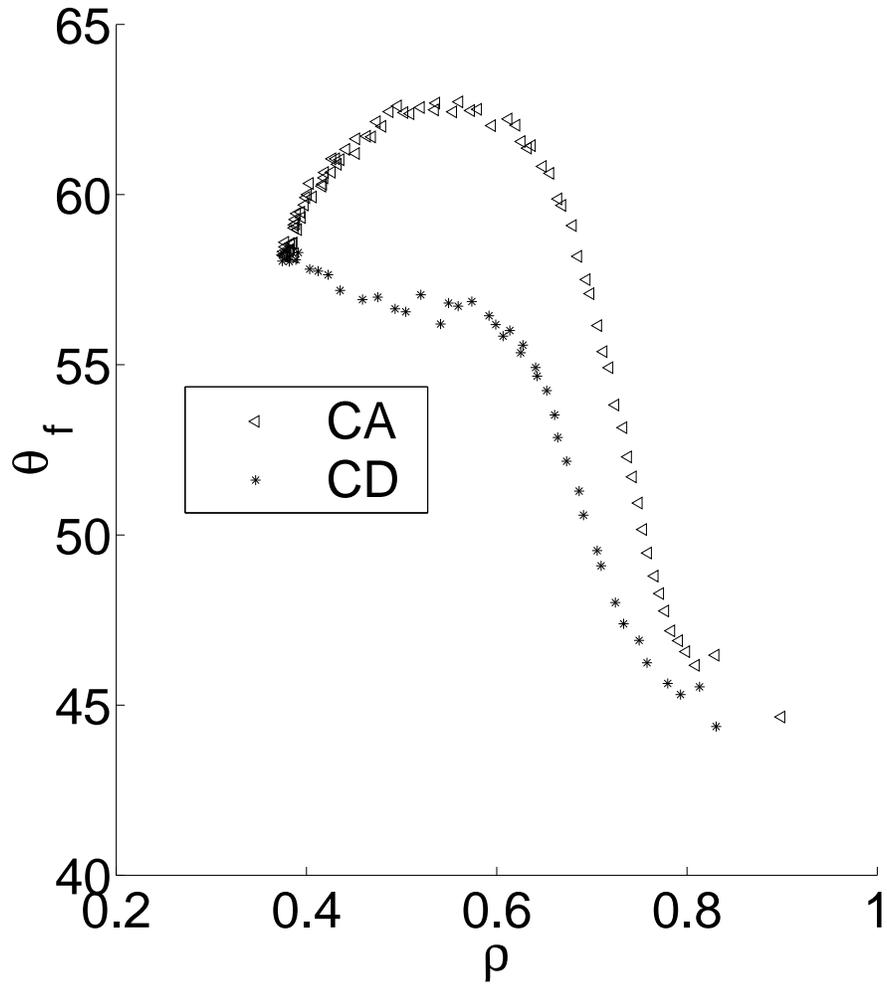} 
\caption{\label{fig13} The major principal directions  
$\theta_f$ (in degrees) of the fabric tensor as a 
function of solid fraction.} 
\end{figure} 

\begin{figure}[h] 
\includegraphics[width=6cm]{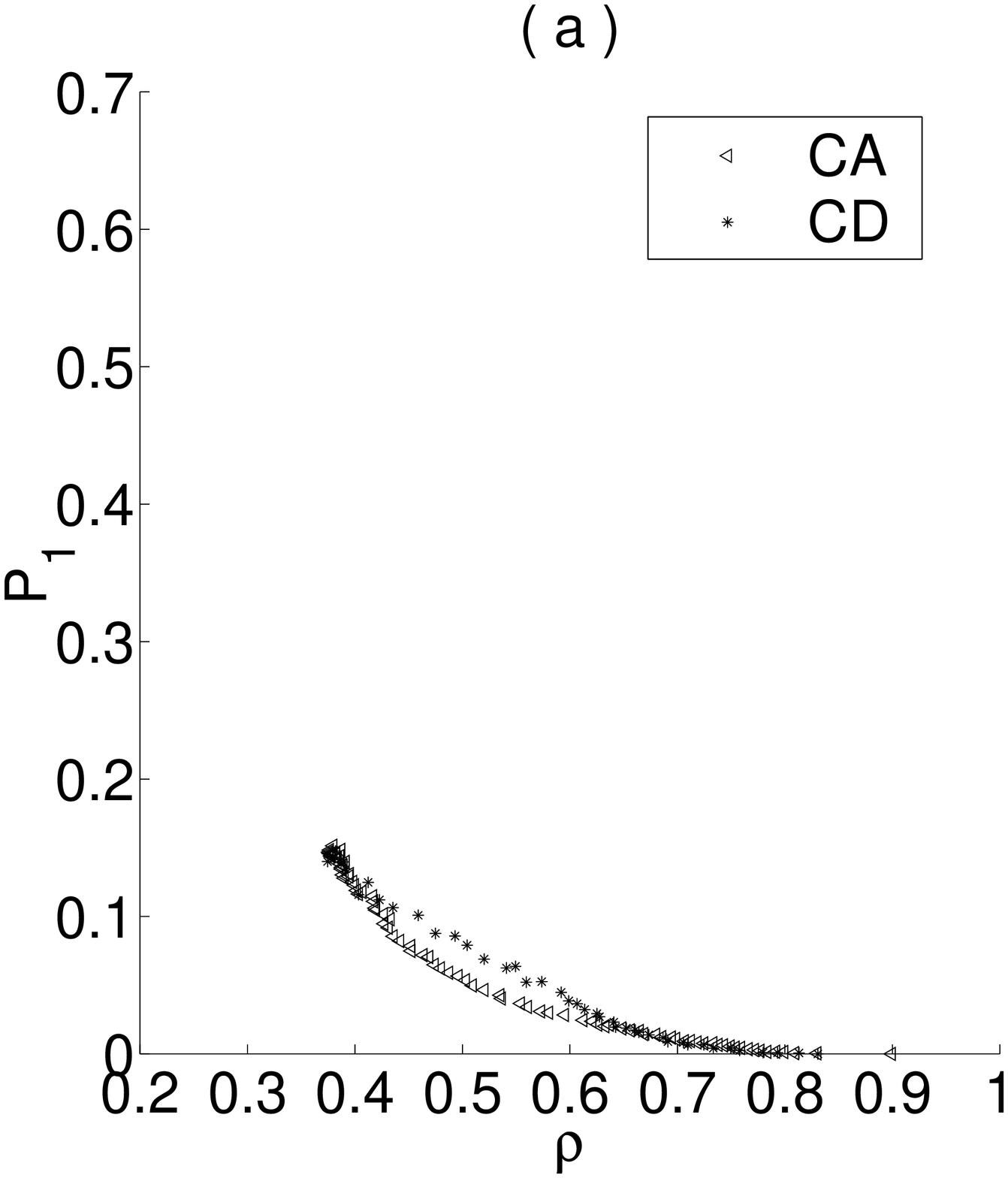} 
\includegraphics[width=6cm]{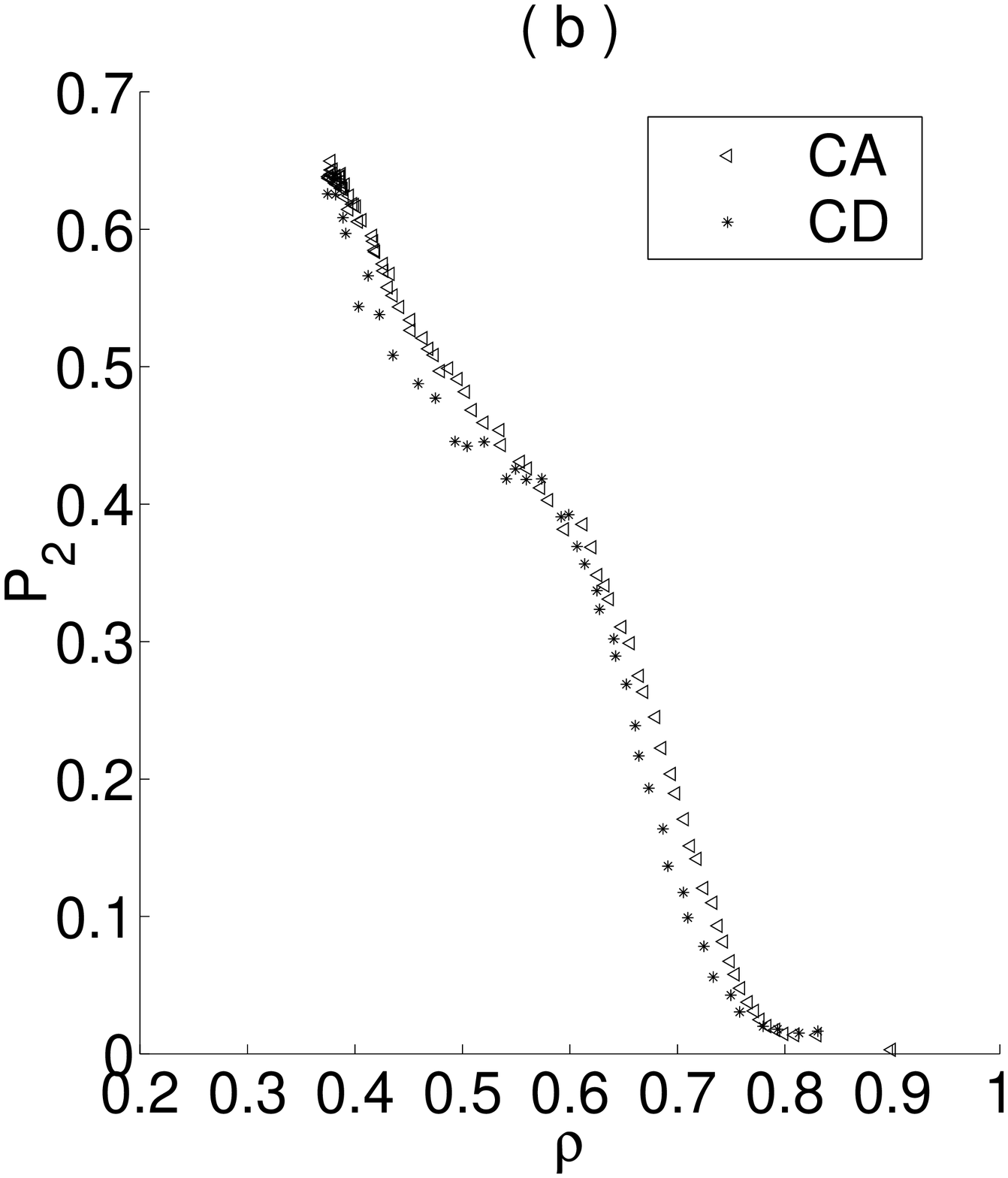} 
\includegraphics[width=6cm]{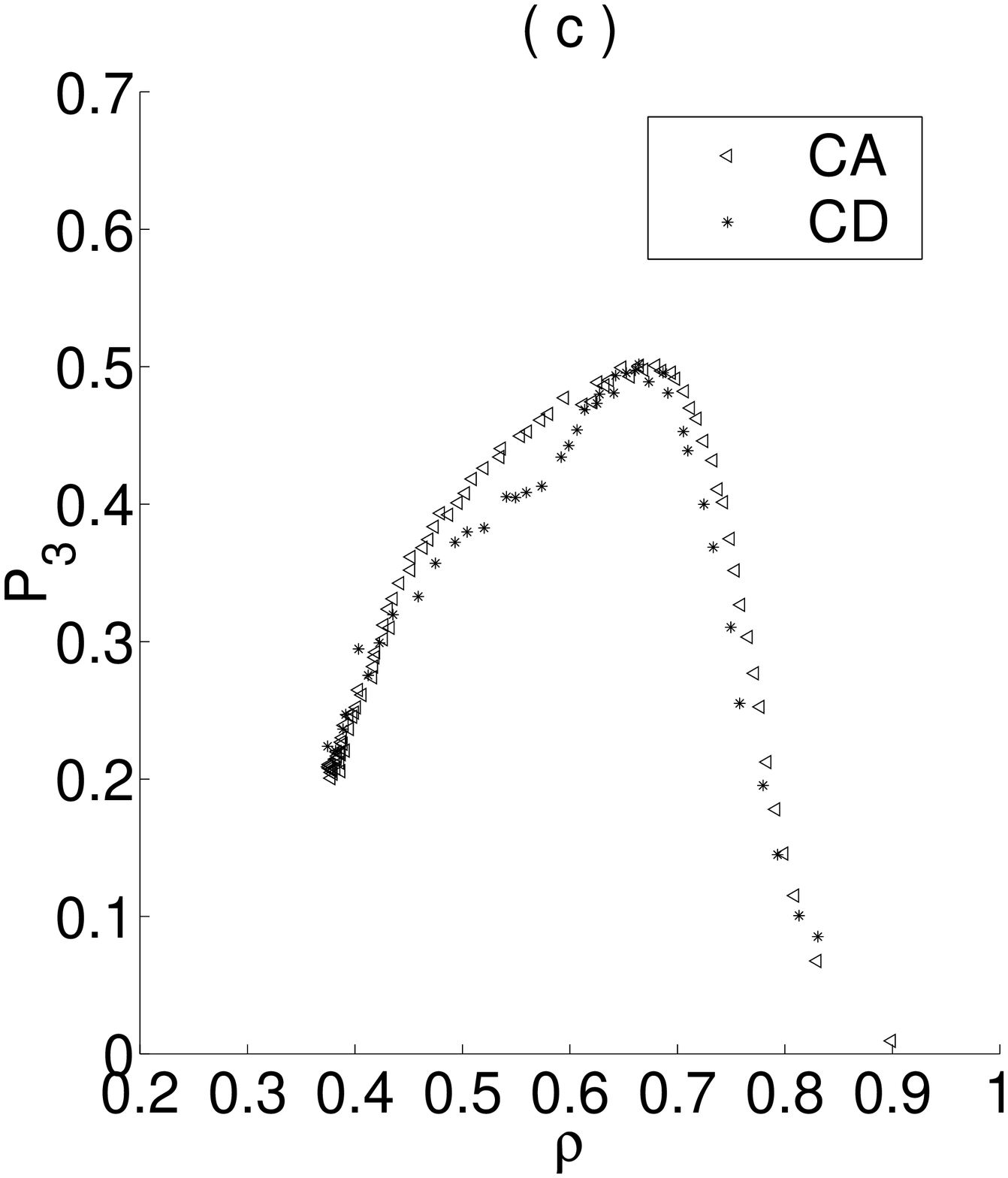} 
\includegraphics[width=6cm]{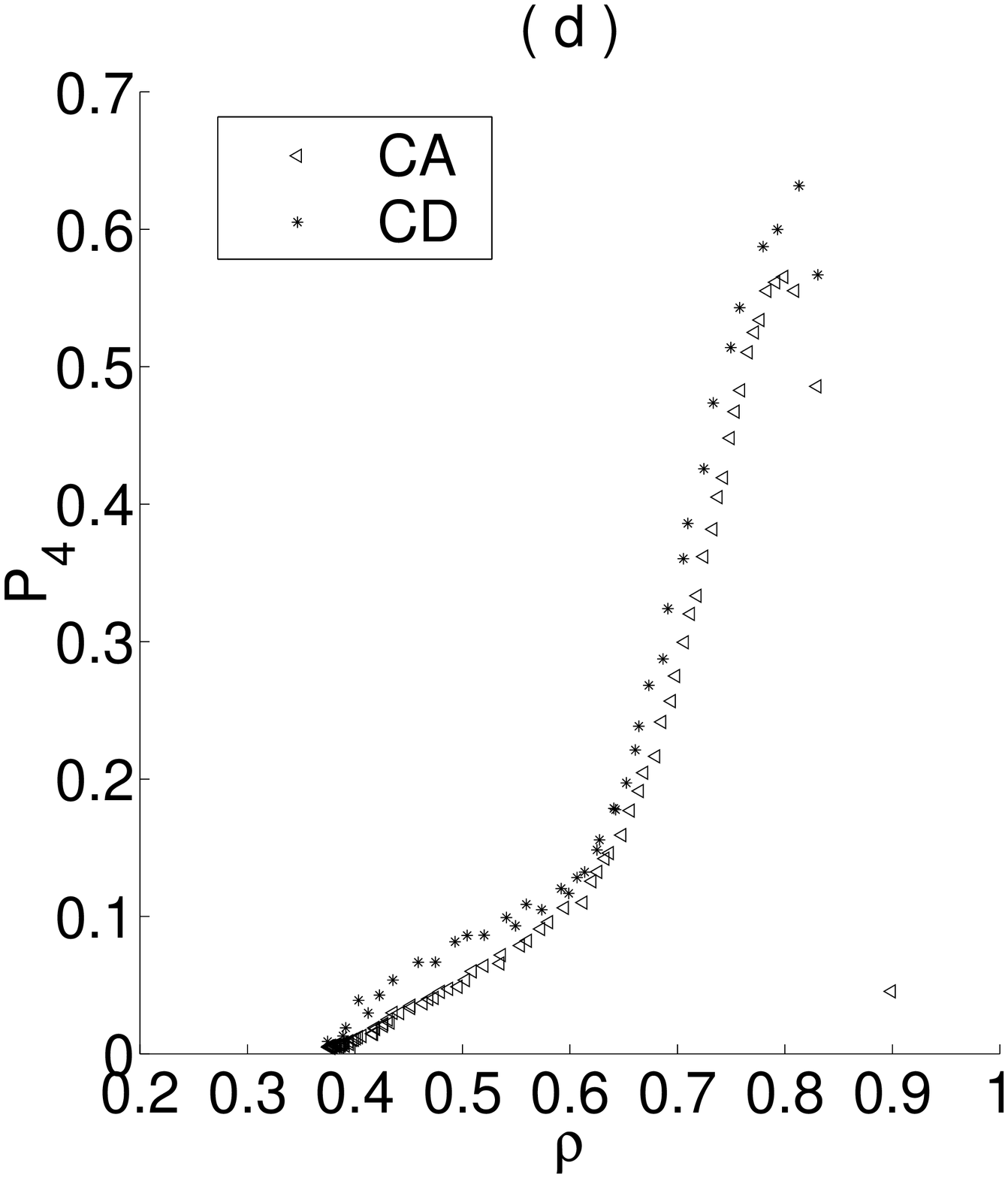} 
\includegraphics[width=6cm]{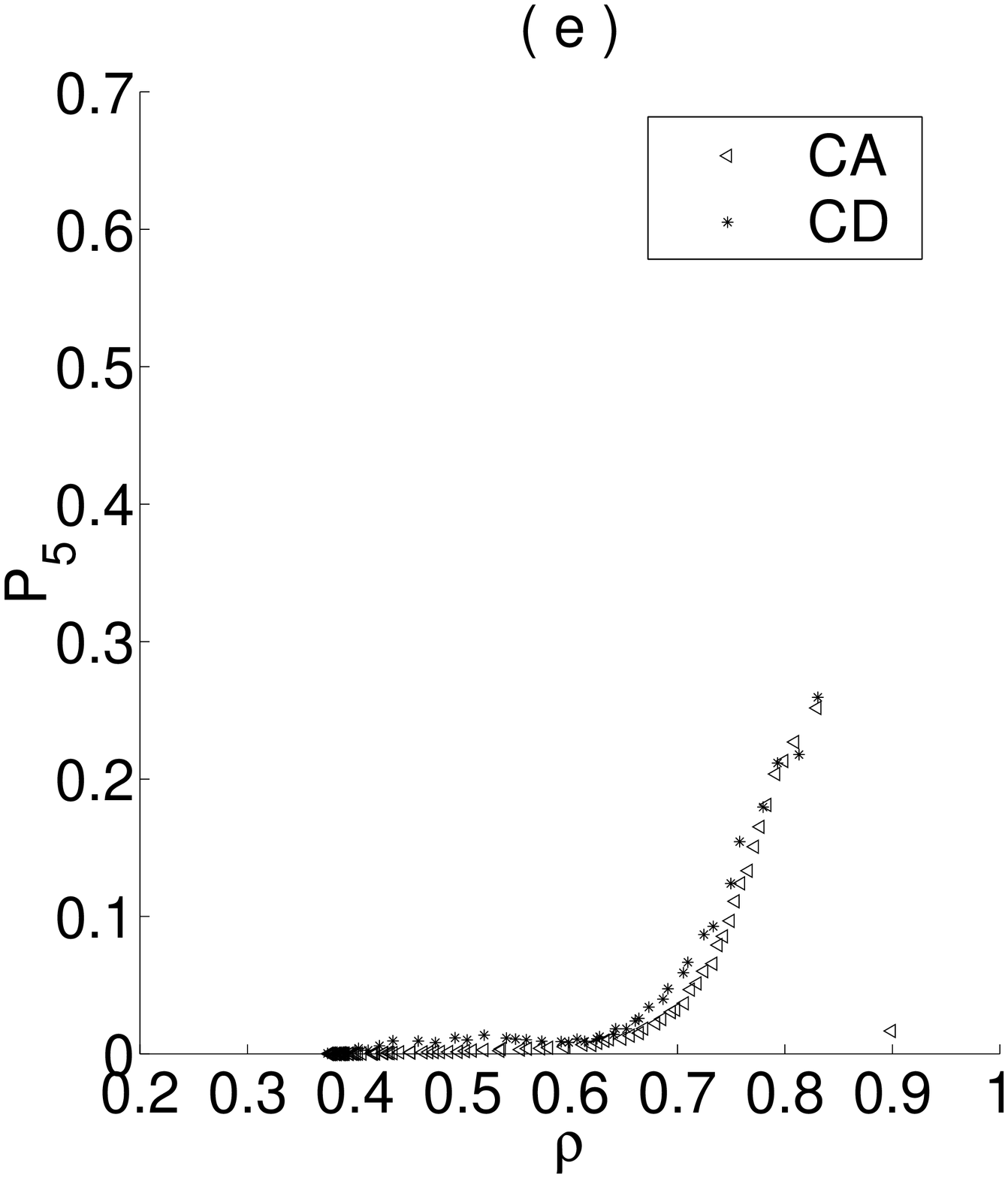} 
\includegraphics[width=6cm]{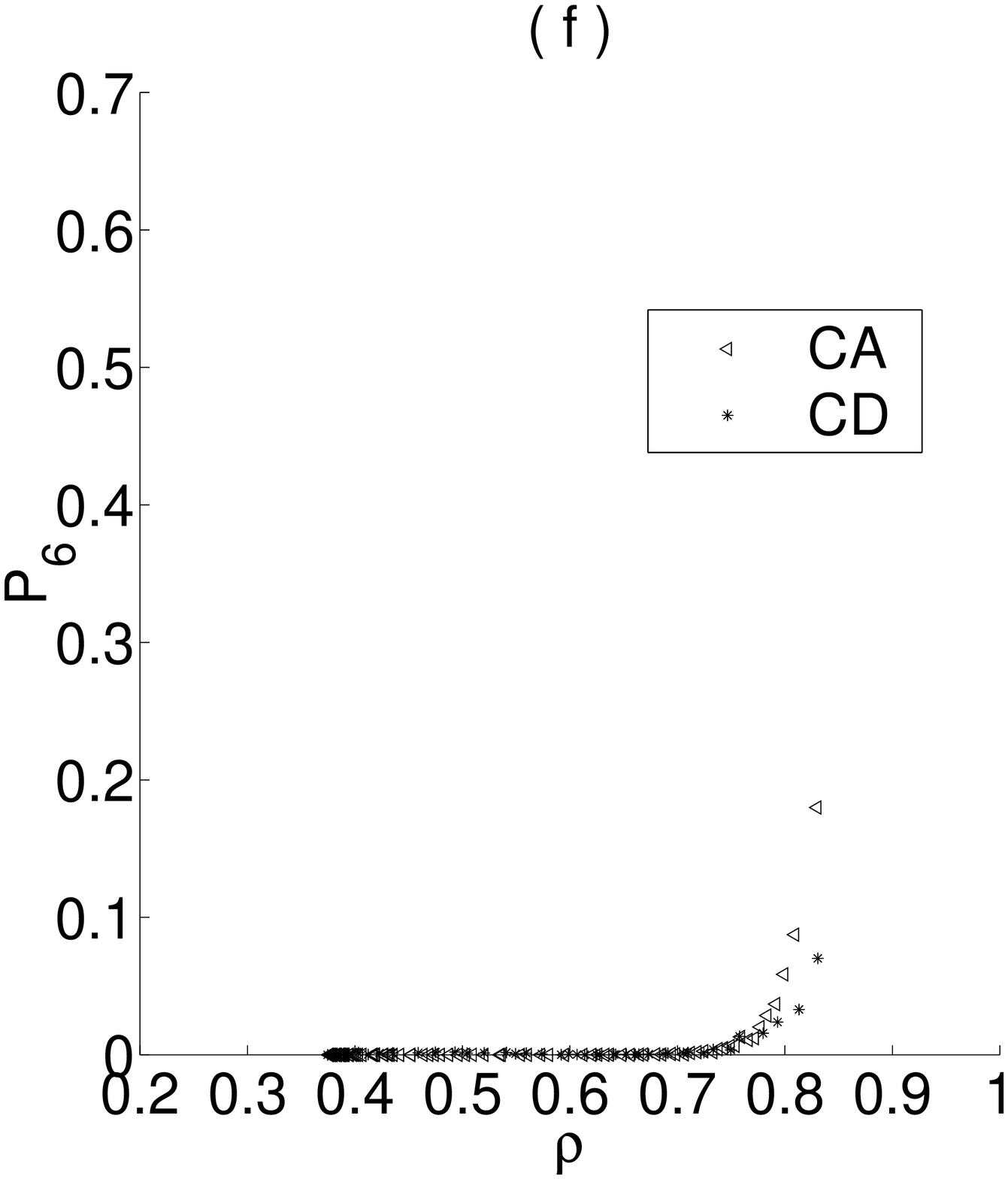} 
\caption{\label{fig14} The connectivity 
numbers $P_1, \cdots, P_6$ as a function of solid fraction.} 
\end{figure} 
\newpage 
\begin{figure}[h] 
\includegraphics[width=8cm]{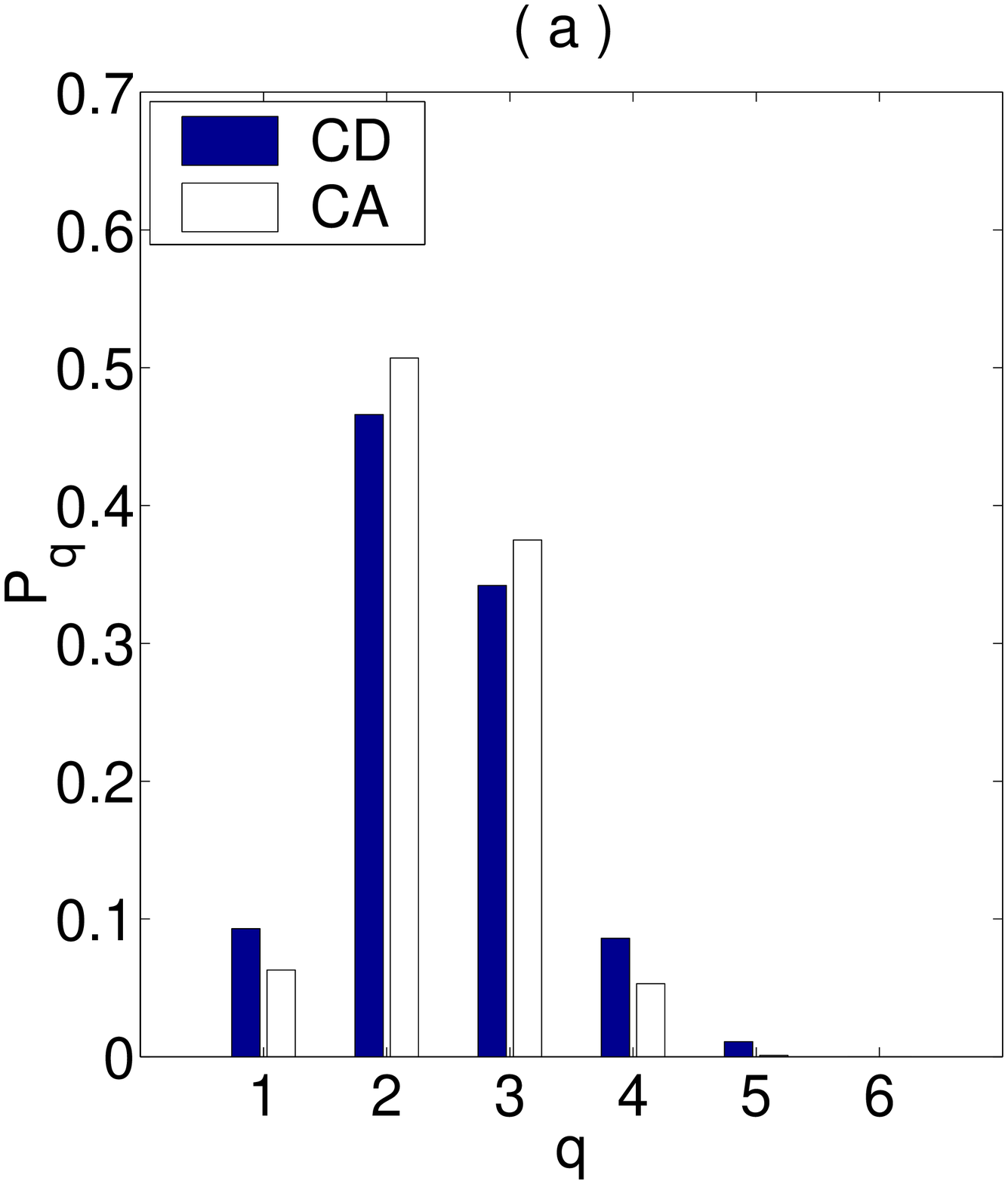} 
\includegraphics[width=8cm]{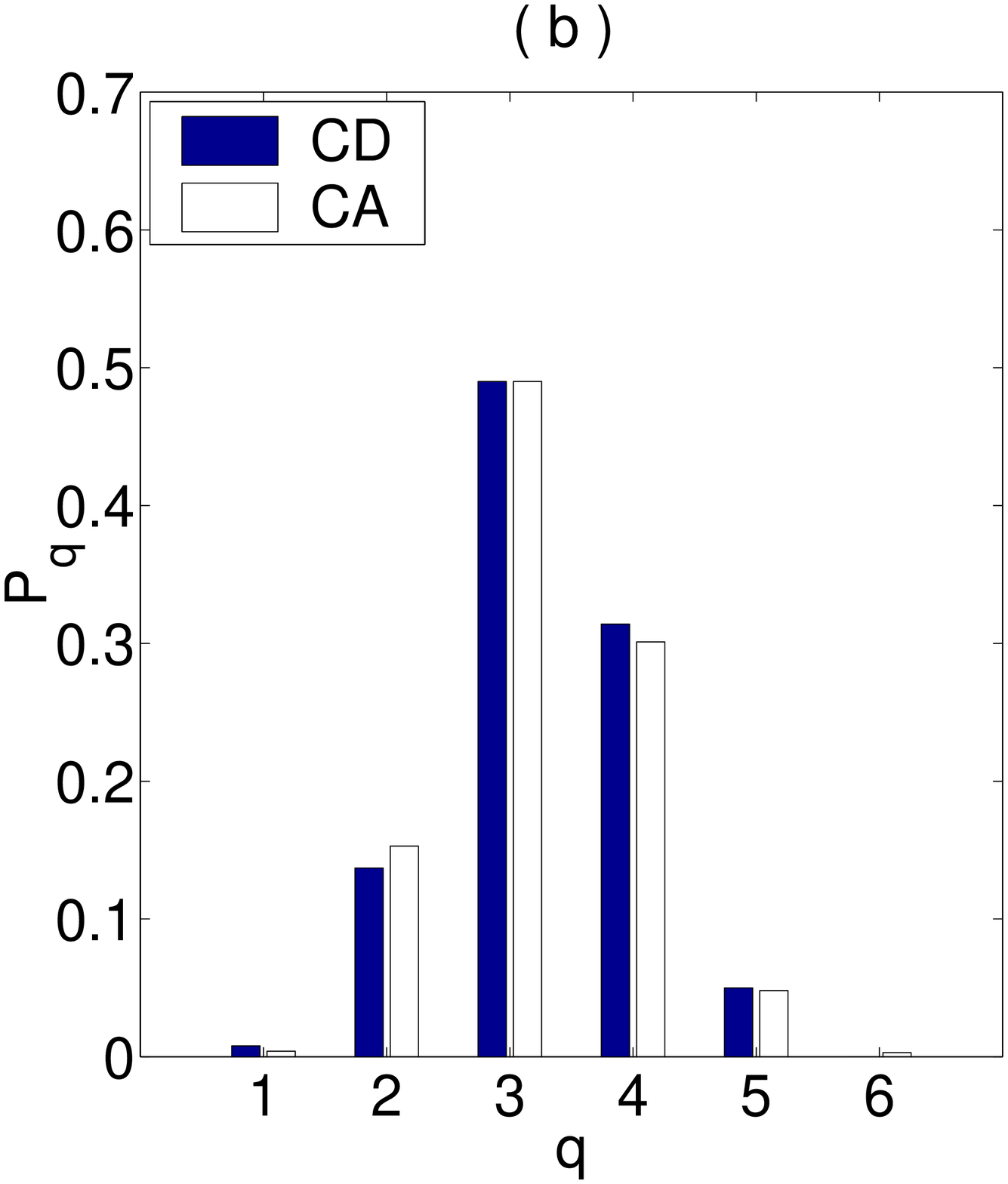} 
\includegraphics[width=8cm]{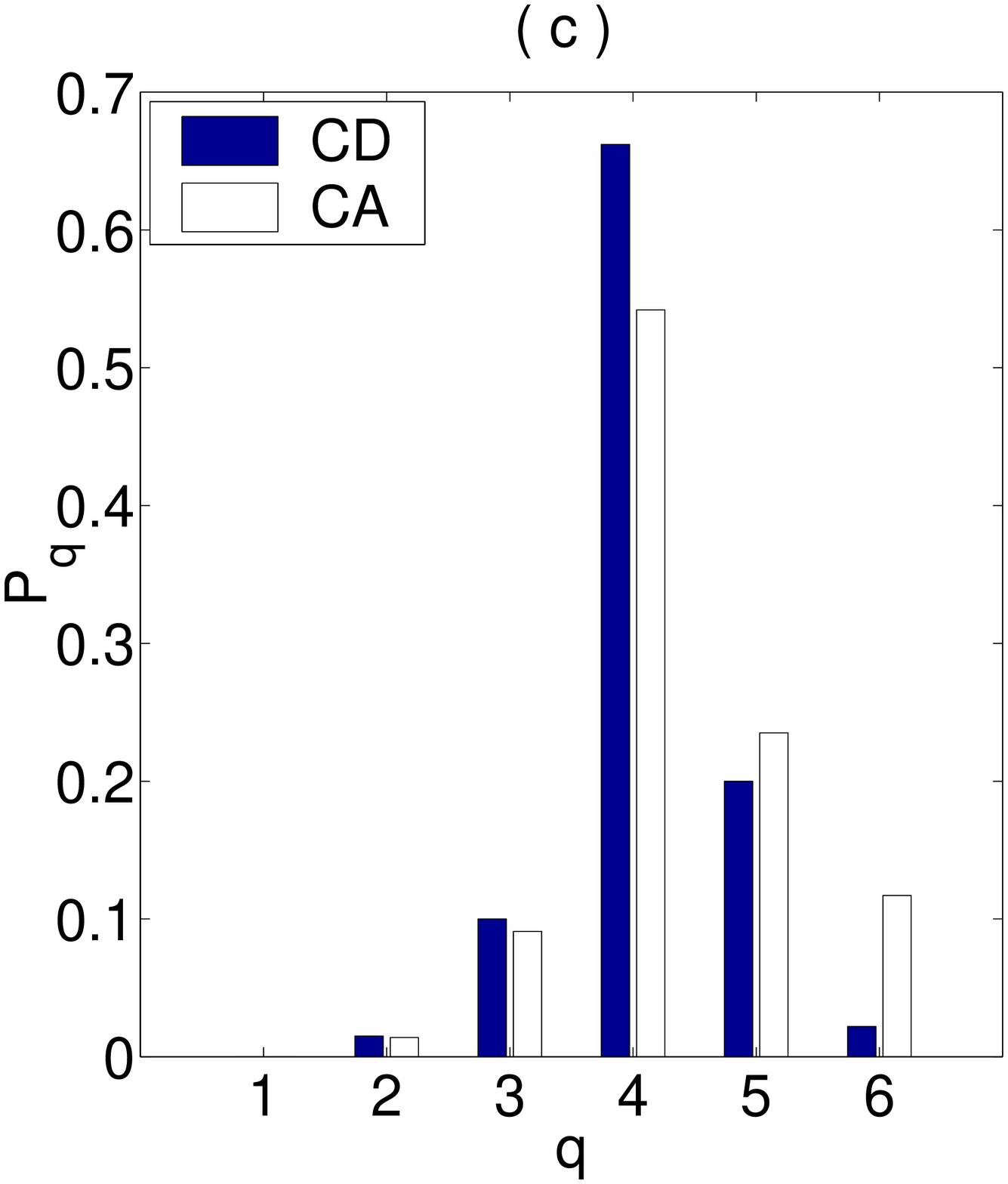} 
\caption{\label{fig15} The connectivity diagram 
of CA and CD packings  packings at three 
different values of solid fraction: (a) $\rho=0.5$, 
(b) $\rho=0.7$ and (c) $\rho=0.8$.} 
\end{figure} 
 
\end{document}